\documentclass[useAMS,usenatbib]{mnras}
\usepackage{graphicx}
\usepackage{color}
\usepackage{amssymb}
\usepackage{amsmath}
\usepackage{subfigure}
\usepackage{rotating}
\usepackage{pdflscape}
\usepackage{bm}
\usepackage{mathtools}
\usepackage[colorinlistoftodos]{todonotes}
\usepackage{ulem}

\usepackage{comment}
\includecomment{details}
\graphicspath{{images/}}

\newcommand\psj{PSJ} 

\title[{Energetic particle environment of GJ\,436}]{The energetic particle environment of a GJ\,436\,b-like planet}
\author[Rodgers-Lee et al.]{D. Rodgers-Lee$^{1,2}$\thanks{E-mail:
dlee@cp.dias.ie}, {P. B. Rimmer}$^{3}$, {A. A. Vidotto}$^{4}$, {A. J. Louca}$^{4}$, {A. M. Taylor}$^{5}$, {A. L. Mesquita}$^{4}$, \newauthor {Y. Miguel}$^{6,4}$, {O. Venot}$^{7}$, {Ch. Helling}$^{8,9}$, {P. Barth}$^{8,10,11,12}$ and {E. Lacy}$^{1,13}$ \\ 
$^{1}$ Dublin Institute for Advanced Studies, School of Cosmic Physics, 31 Fitzwilliam Place, Dublin 2, D02 XF86, Ireland \\
$^{2}$ Trinity College Dublin, School of Physics, University of Dublin, College Green, Dublin 2, D02 PN40, Ireland  \\
$^{3}$ Cavendish Laboratory, University of Cambridge, JJ Thomson Ave, Cambridge, CB3 0HE, United Kingdom\\ 
$^{4}$ Leiden Observatory, Leiden University, P.O. Box 9513, 2300 RA Leiden, The Netherlands \\
$^{5}$ Deutsches Elektronen-Synchrotron, Zeuthen, Germany\\
$^{6}$ SRON, Netherlands Institute for Space Research, Niels Bohrweg 4, 2333 CA Leiden, the Netherlands \\
$^{7}$ Universit\'{e} Paris Cit\'{e} and Univ Paris Est Creteil, CNRS, LISA, F-75013 Paris, France \\
$^{8}$ Space Research Institute, Austrian Academy of Sciences, Schmiedlstrasse 6, A-8042 Graz, Austria \\
$^{9}$ Institute for Theoretical Physics and Computational Physics, Graz University of Technology, Petersgasse 16
8010 Graz \\
$^{10}$ St Andrews Centre for Exoplanet Science, University of St Andrews, North Haugh, St Andrews, KY16 9SS, UK \\
$^{11}$ SUPA, School of Physics and Astronomy, University of St Andrews, North Haugh, St Andrews, KY16 9SS, UK \\
$^{12}$ School of Earth \& Environmental Sciences, University of St Andrews, Bute Building, Queen's Terrace, St Andrews, KY16 9TS, UK \\
$^{13}$ School of Physics, University College Dublin, Belfield, Dublin 4, Ireland \\
}
\definecolor{dg}{rgb}{0.0, 0.6, 0.1}

\begin{document}
\date{Accepted xxxx xxxxxx xx. Received xxxx xxxxxx xx; in original form xxxx xxx xx}
\pagerange{\pageref{firstpage}--\pageref{lastpage}} \pubyear{xxxx}
\maketitle

\label{firstpage}

\begin{abstract}
A key first step to constrain the impact of energetic particles in exoplanet atmospheres is to detect the chemical signature of ionisation due to stellar energetic particles and Galactic cosmic rays. We focus on GJ\,436, a well-studied M dwarf with a warm Neptune-like exoplanet. We demonstrate how the maximum stellar energetic particle momentum can be estimated from the stellar X-ray luminosity. We model energetic particle transport through the atmosphere of a hypothetical exoplanet at orbital distances between $a=0.01-0.2\,$au from GJ\,436, including GJ\,436\,b's orbital distance (0.028\,au). For these distances we find that, at top-of-atmosphere, stellar energetic particles ionise molecular hydrogen at a rate of $\zeta_{\rm StEP,H_2} \sim 4\times10^{-10}-2\times10^{-13}\,\mathrm{s^{-1}}$. In comparison, Galactic cosmic rays alone lead to $\zeta_{\rm GCR, H_2}\sim2\times 10^{-20}-10^{-18} \,\mathrm{s^{-1}}$. At 10au we find that ionisation due to Galactic cosmic rays equals that of stellar energetic particles: $\zeta_{\rm GCR,H_2} = \zeta_{\rm StEP,H_2} \sim 7\times10^{-18}\,\rm{s^{-1}}$ for the top-of-atmosphere ionisation rate. At GJ\,436\,b's orbital distance, the maximum ion-pair production rate due to stellar energetic particles occurs at pressure $P\sim 10^{-3}\,$bar while Galactic cosmic rays dominate for $P>10^2\,$bar. These high pressures are similar to what is expected for a post-impact early Earth atmosphere. The results presented here will be used to quantify the chemical signatures of energetic particles in warm Neptune-like atmospheres.
\end{abstract}

\begin{keywords}
planetary systems -- planets and satellites: atmospheres -- cosmic rays -- methods: numerical -- stars: low-mass -- stars: winds, outflows
\end{keywords}

\section{Introduction}
\label{sec:intro}

Spectroscopic observations with JWST \citep{gardner_2006,rigby_2022} and future dedicated exoplanet missions, such as Ariel \citep{tinetti_2021}, will characterise the composition of exoplanet atmospheres. Chemical models have predicted that energetic particles lead to specific chemical signatures in exoplanet atmospheres, such as $\rm{H_3O^+}$ for gas giants \citep{helling_2019,barth_2021}. \citet{bourgalais_2020} produced synthetic JWST and Ariel transmission spectra of a sub-Neptune and showed that absorption due to $\rm{H_3O^+}$ should be observable. The detection of such a chemical signature would be a key step to constrain the energetic particle fluxes impacting on planets outside the solar system. Here, we focus on modelling the fluxes of two types of energetic particles: stellar energetic particles from the host star (also known as stellar cosmic rays) and Galactic cosmic rays from the interstellar medium (ISM).   In the future, it may be possible to disentangle the effect of stellar energetic particles and Galactic cosmic rays by detecting chemical signatures of energetic particles for exoplanets at different orbital distances \citep[as suggested in][for instance]{rodgers-lee_2020b}.

Galactic cosmic rays interact with, and their fluxes are suppressed by, the magnetised winds of low-mass stars (e.g. \citet{potgieter_2013} for the Sun and \citet{rodgers-lee_2021b} for solar-like stars). Previous studies have focused on calculating the Galactic cosmic ray fluxes in the habitable zone \citep[i.e. the orbital distances where liquid water can exist,][]{kasting_1993,pierrehumbert_2011,abe_2011,kopparapu_2013,zsom_2013} and at the orbital distance of known exoplanets for a number of well-studied solar-type \citep{rodgers-lee_2021b} and M dwarf stars \citep{herbst_2020,mesquita_2021,mesquita_2022a,mesquita_2022b}. 

Here, our aim is to combine the energetic particle transport through the stellar system with the subsequent transport through the exoplanet atmosphere. We relate the maximum energy of the stellar energetic particles accelerated by stellar flares to stellar magnetic field strength, following \citet{rodgers-lee_2021a}, and show how this can be related to stellar X-ray luminosity. Previous studies  have investigated the distribution of stellar energetic particles in the Trappist-1 and AU Mic systems \citep{fraschetti_2019,fraschetti_2022}. In the latter case, following a perturbation of the interplanetary medium by a coronal mass ejection (CME). Separately, the propagation, and chemical effect, of energetic particles through a number of Earth-like \citep[e.g.][]{griessmeier_2015,tabataba-vakili_2016,herbst_2019,scheucher_2020} and hot Jupiter/brown dwarf \citep{helling_2019,barth_2021} atmospheres has been studied. 

Here, we focus on the environment of the warm Neptune exoplanet GJ\,436\,b. GJ\,436\,b is scheduled for JWST observations and is also an Ariel target \citep{edwards_2022}. We calculate the ionisation rate, ion-pair production rate and skin-depth equivalent dose rate due to energetic particles at different heights in the exoplanet atmosphere. In addition to studying GJ\,436\,b, we also investigate the effect of artificially changing the exoplanet's orbital distance on these quantities. The energetic particle ionisation rate is an important input for studying disequilibrium chemistry in exoplanet atmospheres, with chemical kinetic models such as ARGO \citep{rimmer_2016,rimmer_2019b}. Here, for simplicity, we do not account for the effect of planetary magnetic fields. However, planetary magnetic fields will reduce the energetic particle fluxes penetrating the exoplanet atmosphere \citep{herbst_2013,herbst_2019,griessmeier_2015,griessmeier_2016}. 

This study has been separated into two companion papers. This paper (Paper I) focuses on characterising the energetic particle environment of a GJ\,436\,b-like planet. The paper is structured as follows: Section\,\ref{subsec:cr_fluxes} introduces the models that are used for the energetic particle transport and in Section\,\ref{sec:results} our results are presented. The discussion and conclusions are given in Sections\,\ref{sec:discussion} and \ref{sec:conclusions}. Paper II (\citealt{Rimmer2023}) will focus on the chemical consequences of the energetic particles for this system.

\section{Energetic particle spectra}
\label{subsec:cr_fluxes}

\begin{figure*}
	\centering\includegraphics[width=\textwidth]{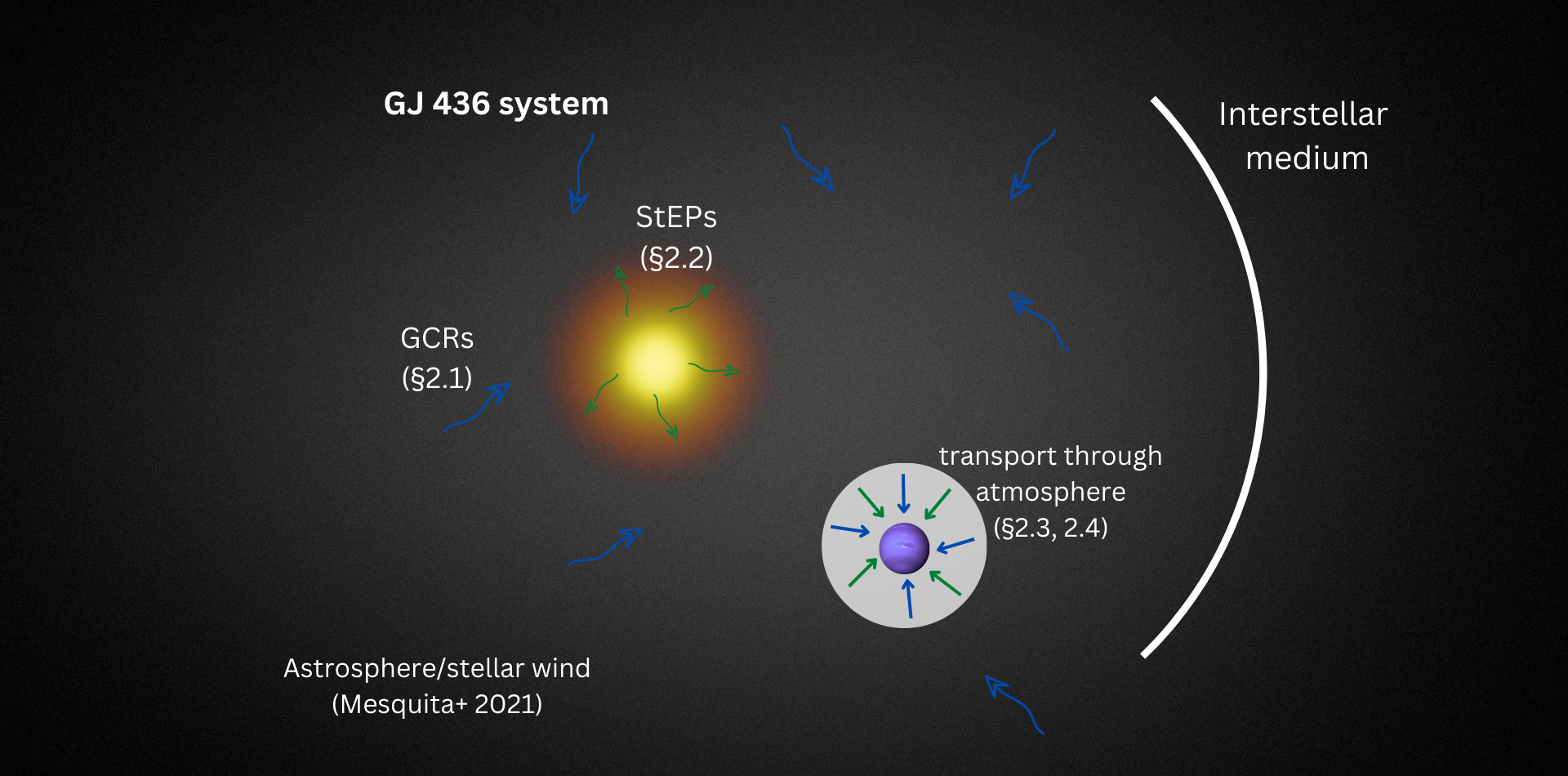}
       	\centering
  \caption{Schematic of the GJ\,436 system (not to scale) and its surrounding environment embedded in the ISM. The curved blue and green arrows represent the advection and diffusion of Galactic cosmic rays (labelled as `GCRs') and stellar energetic particles (labelled as `StEPs'), respectively. The straight blue and green arrows represent their ballistic transport through the exoplanet atmosphere which does not account for the effect of a planetary magnetic field. The annotations indicate where the stellar wind model and the various energetic particle transport models are described or were first presented. It is important to note, as indicated in the figure, that the stellar energetic particle and Galactic cosmic ray fluxes are assumed to be isotropic at the top of the atmosphere. This assumption is discussed further in Section\,\ref{subsec:cr_atmos}.\label{fig:sketch}}
\end{figure*}

We consider the transport of both Galactic cosmic rays from the ISM and stellar energetic particles, described in Sections \ref{subsec:gcr_fluxes} and \ref{subsec:step_fluxes}. In this section we also describe our method to construct the stellar energetic particle spectra. Fig.\,\ref{fig:sketch} shows a schematic of the system indicating the two sources of energetic particles. The exoplanet atmosphere profiles that we use are described in Section\,\ref{subsec:atmos_profile}. Using these, we then model the energetic particle propagation through the exoplanet atmosphere, which is described in Section\,\ref{subsec:cr_atmos}. 

\subsection{Galactic cosmic ray fluxes}
\label{subsec:gcr_fluxes}
The Galactic cosmic ray fluxes throughout the GJ\,436 system were previously calculated \citep{mesquita_2021} using a diffusion-advection cosmic ray transport model \citep[described in][]{rodgers-lee_2020b}. The stellar wind velocity and magnetic field profile are important quantities to model the cosmic ray transport through the stellar system. The Galactic cosmic ray fluxes that we adopt here correspond to `Case A' from \citet{mesquita_2021}, using the stellar wind properties from \citet{mesquita_2020}. Fig.\,\ref{fig:gcr_fluxes} shows the differential intensity, $j$ (i.e. the number of energetic particles per unit area, steradian, time and kinetic energy), as a function of energetic particle kinetic energy for the Galactic cosmic rays. The black line represents the Galactic cosmic ray fluxes at the orbital distance of GJ\,436\,b ($\sim0.03\,$au). The coloured lines represent $j$ at the different orbital distances between $a=0.01-0.2\,$au. For the temperature pressure profiles in Section\,\ref{subsec:atmos_profile}, we focus on a range of orbital distances between $a=0.01-0.2\,$au in increments of 0.01\,au. We select the Galactic cosmic ray fluxes from the cosmic ray transport model closest to these orbital distances. However, the cosmic ray transport uses a logarithmic grid and due to the model resolution some of the lines overlap. This occurs at larger orbital distances, for $a>0.11$\,au in all but one case, where the fluxes are already very similar (see Fig.\,\ref{fig:gcr_fluxes}). Thus, the overlap will not affect our results significantly. The Galactic cosmic ray fluxes vary significantly for $\lesssim 10\,$GeV energy cosmic rays depending on the orbital distance considered. The fluxes for $\lesssim 10\,$GeV energy cosmic rays are also much smaller than the values observed at Earth \citep[see Fig.5 from][]{mesquita_2021} at all of the orbital distances considered here.

Case A corresponds to a stellar wind model which is more magnetically-dominated than the more thermally-dominated wind \citep[`Case B’ in][]{mesquita_2021}. Fig.\,2 from \citet{mesquita_2021} shows the stellar wind velocity and magnetic field profiles for Case A and B. For Case A, the surface stellar magnetic field adopted was $B_\star=4\,$G, the stellar wind terminal velocity, $v_\infty$, was found to be 1250\,${\rm km\,s^{-1}}$ and the stellar mass loss rate was $\dot{M_\star}=1.2\times 10^{-15}M_\odot\,{\rm yr^{-1}}$. The astrospheric size ($R_\mathrm{ast}$), which determines how far the Galactic cosmic rays must travel through the stellar wind, was calculated using $v_\infty$ and $\dot{M_\star}$ to be 33\,au \citep[see Eq.\,11 in][]{mesquita_2021}.

Another important quantity for the transport model is the assumed Galactic cosmic ray spectrum outside of the astrosphere. A model fit to the local interstellar spectrum (LIS) for Galactic cosmic rays \citep{vos_2015} from \textit{Voyager 1} data \citep{stone_2013} was used as the boundary condition for these simulations. Finally, the assumed turbulence properties for the stellar wind are important by dictating the diffusion of the Galactic cosmic rays from the ISM through the astrosphere. For more details see \citet{mesquita_2021}. 

\subsection{Stellar energetic particle fluxes}
\label{subsec:step_fluxes}
Unlike the LIS for Galactic cosmic rays, less is known about stellar energetic particle spectra. It is generally assumed that stars more active than the Sun will produce higher fluxes of stellar energetic particles \citep[e.g.][]{feigelson_2002}. The higher flux is expected due to the increased flare energies and flaring rate of active stars \citep[e.g. from Kepler and TESS observations,][]{maehara_2015,gunther_2020}.

It is also very likely for active stars that stellar energetic particles will be accelerated to higher energies than for the Sun. The Sun itself has been inferred to accelerate particles to GeV energies from $\gamma-$ray observations \citep{ajello_2014,ackermann_2014}. Considering the maximum energy that stellar energetic particles are accelerated to by their host star is important because the energy loss rate for energetic particles is energy dependent. The energy loss rate for GeV energy particles is much lower than for $\sim$MeV energy particles \citep[see Fig.\,5.6 in][]{longair_2011}. Thus, GeV energy particles are far more penetrating than $\sim$MeV energy particles and can, for instance, lead to showers of secondary particles that can reach the surface of a rocky exoplanet \citep{atri_2017}.

Following \citet{rodgers-lee_2021a}, there are three quantities that we use to obtain a stellar energetic particle spectrum. These are (i) the power law index for the spectrum, (ii) the total energy available to produce stellar energetic particles and (iii) the maximum stellar energetic particle momentum ($p_{\rm max}$) before the spectrum decays exponentially. The power law index is set to be $\alpha=2$ \citep[representative of diffusion shock acceleration,][and compatible with acceleration due to magnetic reconnection]{krymskii_1977,bell_1978,blandford_1978} such that $dN/dp \propto p^{-\alpha}e^{-p/p_{\rm max}}$, where $N$ and $p$ are the number and momentum of the particles, respectively. The total luminosity injected in stellar energetic particles ($L_\mathrm{CR}$) that we adopt is $0.1P_\mathrm{SW}$, where $P_\mathrm{SW}=\dot{M_\star}v_\infty^2/2$ is the stellar wind kinetic power. This is in line with efficiency estimates from supernova remnants \citep{vink_2010}. We calculate $L_\mathrm{CR}=6\times 10^{25}{\rm erg\,s^{-1}}$ for GJ\,436 with $\dot{M_\star}=1.2\times 10^{-15}M_\odot\,{\rm yr^{-1}}$ and $v_\infty=1250\,{\rm km\,s^{-1}}$.

\subsubsection{Maximum stellar energetic particle momentum}
In \citet{rodgers-lee_2021a}, the Hillas criterion \citep{hillas_1984} was used to estimate the maximum stellar energetic particle momentum that the Sun would have accelerated particles to in the past. The Hillas criterion posits that the maximum energy achieved by an accelerator is limited by the accelerator size, i.e. if the accelerated particle's Larmor radius is larger than the accelerator region it will escape. Using the Hillas criterion, the maximum momentum, $p_\mathrm{max}$, of accelerated particles from a star can be expressed as 
\begin{equation}
\frac{p_\mathrm{max}}{p_\mathrm{max,\odot}} = \left( \frac{\beta_\star B_\star R_\star}{\beta_\odot B_\odot R_\odot} \right)
\label{eq:hillas}
\end{equation}
\begin{figure*}%
	\centering
    \subfigure[]{%
        \includegraphics[width=0.5\textwidth]{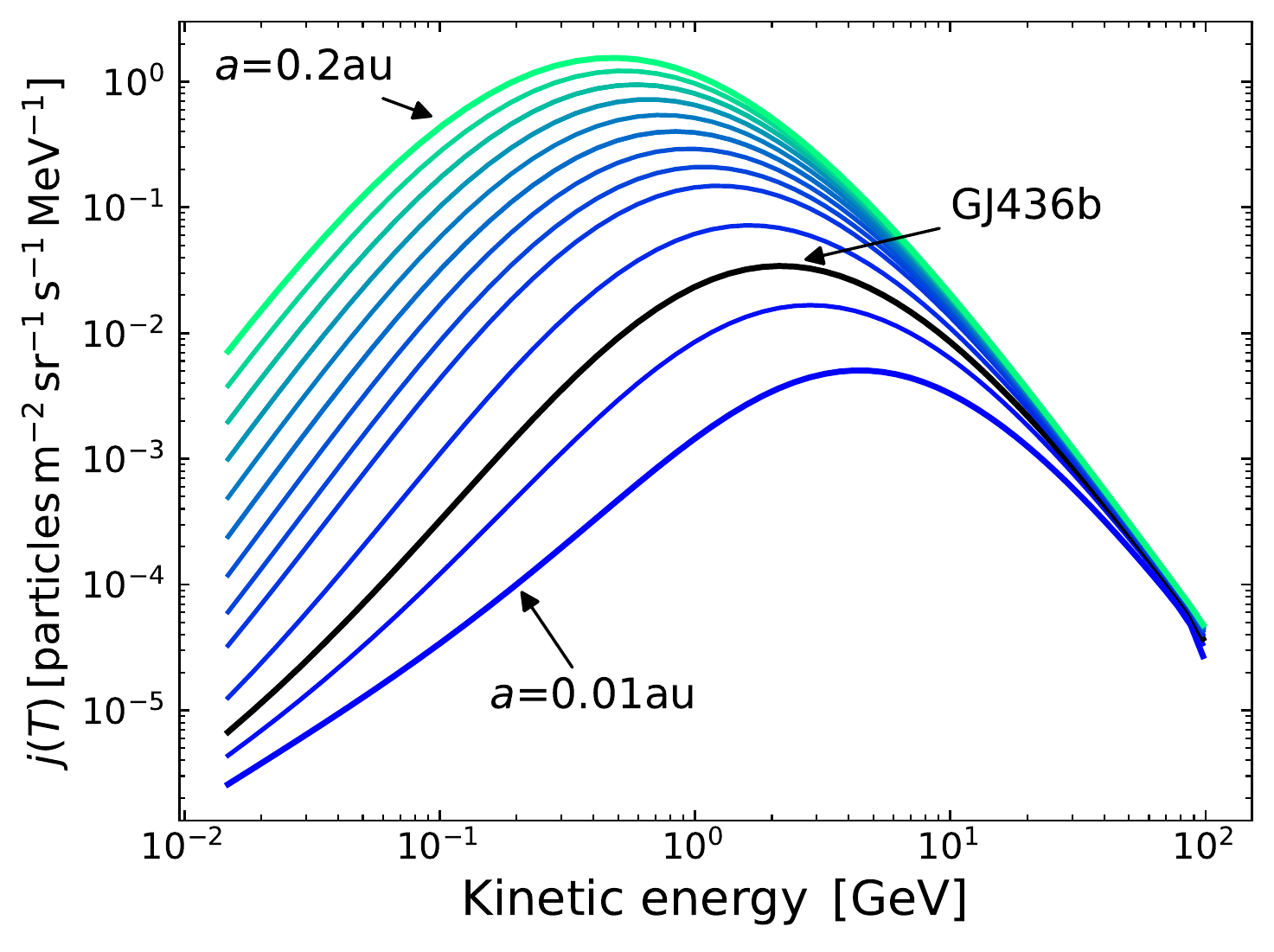}
       	\centering
\label{fig:gcr_fluxes}}%
~
	\centering
    \subfigure[]{%
        \includegraphics[width=0.5\textwidth]{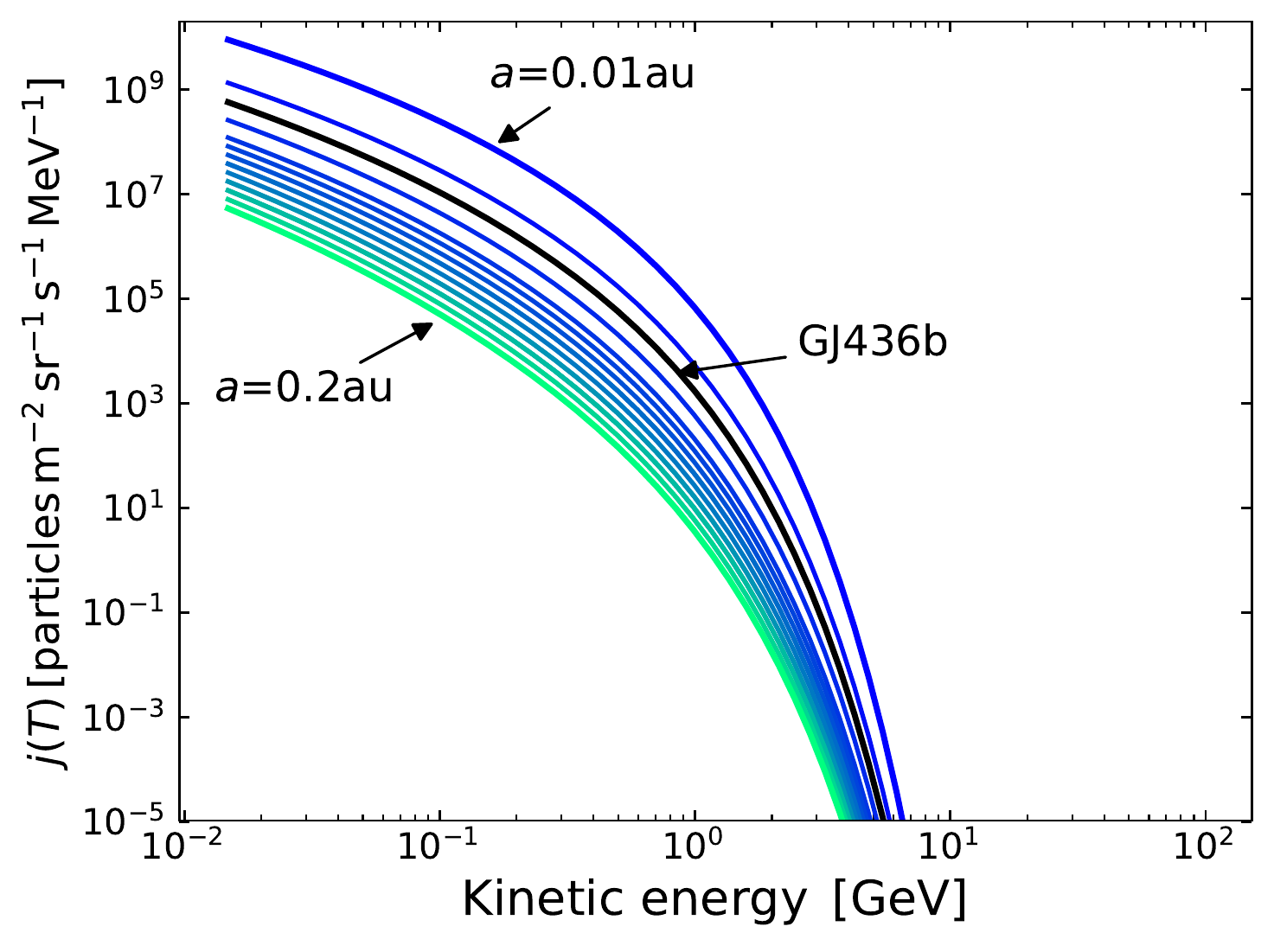}
       	\centering
\label{fig:stcr_fluxes}}%
    \caption{The (a) Galactic cosmic ray and (b) stellar energetic particle differential intensities, $j$, are plotted as a function of the energetic particle kinetic energy at different orbital distances. The colour of the lines represents the orbital distance. For instance, for the left hand plot, the lowest blue line represents the fluxes at $a=0.01\,$au and the highest green line at $a=0.2\,$au. Additionally, the solid black line represents the fluxes at the orbital distance of GJ\,436\,b. The Galactic cosmic ray fluxes are from \citet{mesquita_2021}.} 
    \label{fig:cr_fluxes}%
\end{figure*}
where $R_\odot$ and $R_\star$ are the solar and stellar radii, respectively.   $B_\odot$ and $B_\star$ are the surface solar and stellar magnetic field strengths, respectively. $B_\odot=1.3\,$G and $p_\mathrm{max,\odot}=0.2\,$GeV$/c$. The parameters $\beta_\odot$ and $\beta_\star$ are the solar and stellar shock velocity in units of the speed of light, respectively. Here, we assume $\beta_\star\sim \beta_\odot$ as a first approximation. We consider flare accelerated particles injected at $\sim1.4R_\star$ rather than particles accelerated by CMEs \citep[see][for a discussion of this]{rodgers-lee_2021a}. Eq.\,\ref{eq:hillas} is the same as Eq.\,7 from \citet{rodgers-lee_2021a} for the Sun. The only difference is that $R_\star$ is also included in Eq.\,\ref{eq:hillas}.

The Zeeman Doppler Imaging (ZDI) technique has provided average large-scale stellar magnetic field strengths for hundreds of low-mass stars \citep[e.g.][]{donati_2006,morin_2012,moutou_2017}. This technique has traditionally focused on active stars which generally have strong magnetic fields, best suited for the observations. However, transiting exoplanets are typically detected around less active stars because increased stellar activity, in the form of stellar spots, tends to obscure exoplanet signatures. Thus, the overlap between these samples is currently extremely limited and stellar magnetic field strengths for less active stars often must be estimated. For instance, \citet{vidotto_2014} presented the relation between stellar X-ray luminosity ($L_{\rm X}$) and the large-scale magnetic flux ($\Phi$) derived from a large sample of stars with ZDI maps, as 
\begin{equation}
L_{\rm X} \approx 10^{-13.7}\Phi^{1.80\pm 0.20} \mathrm{erg\,s^{-1}}
\label{eq:lx}
\end{equation}
where $\Phi = 4\pi R_\star^2\langle |B|\rangle$\,(Mx) and $\langle |B|\rangle$ is the average large-scale magnetic field strength derived using the ZDI technique. Using Eq.\,\ref{eq:lx}, we can now re-write Eq.\,\ref{eq:hillas} in terms of $L_{\rm X}$ as
\begin{eqnarray}
\frac{p_\mathrm{max}}{p_\mathrm{max,\odot}} 
  &=& \left( \frac{B_\star R_\star^2}{B_\odot R_\odot^2 } \right) \left(\frac{R_\odot}{R_\star}\right) = \left(\frac{\Phi}{\Phi_{\odot}}\right) \left(\frac{R_\odot}{R_\star}\right) \nonumber \\
  &=& \left( \frac{L_{\rm X,\star}}{L_{\rm X,\odot}}\right)^\gamma\left(\frac{R_\odot}{R_\star}\right). 
  \label{eq:pmax_new}
\end{eqnarray}
where $\gamma = 0.56^{-0.06}_{+0.125}$. Thus, in the absence of a magnetic field strength measurement the maximum stellar energetic particle momentum can be estimated using the stellar X-ray luminosity. Here however, we have used $B_\star=4$\,G for GJ\,436 to be consistent with \citet{mesquita_2021}\footnote{As mentioned in \citet{mesquita_2021}, $B_\star=4$\,G was chosen because the stellar wind model from \citet{mesquita_2020} with this $B_\star$ value predicts an X-ray luminosity that is consistent with what is observed for GJ\,436.}. This results in $p_{\rm max}=0.3$GeV$/c$ using Eq.\,\ref{eq:hillas}. In comparison, using Eq.\,\ref{eq:pmax_new}, we find $p_{\rm max} = 0.1-0.7{\rm GeV}/c$ for $L_{\rm X,\odot} = 2.7\times 10^{26}-4.7\times10^{27}\,{\rm erg\,s^{-1}}$ \citep[spanning the range of values from solar minimum to maximum,][]{peres-2000} with $\gamma=0.56$ and $L_{\rm X,\star}=5.7\times10^{26}{\rm erg\,s^{-1}}$ for GJ\,436 \citep{mesquita_2020}. 

We construct the stellar energetic particle spectrum close to the surface of the star using the values motivated above for $L_\mathrm{CR}$, $\alpha$ and $p_{\rm max}$. This spectrum is used in the energetic particle transport equation \citep[Eq.\,1 from][]{rodgers-lee_2021a} to calculate the resulting differential intensity, $j$, as a function of orbital distance. Fig.\,\ref{fig:stcr_fluxes} shows $j$ as a function of energetic particle kinetic energy for stellar energetic particles at different orbital distances. The black line represents the energetic particle fluxes at the orbital distance of GJ\,436\,b ($\sim0.03\,$au). The coloured lines represent $j$ at different orbital distances between $a=0.01-0.2\,$au in increments of 0.01\,au. The stellar energetic particle fluxes vary significantly for $\lesssim 10\,$GeV energy cosmic rays depending on the orbital distance considered. The stellar energetic particle fluxes are much higher than the Galactic cosmic ray fluxes at MeV energies (see Fig.\,\ref{fig:gcr_fluxes}). For $\gtrsim$GeV energies the Galactic cosmic rays begin to dominate due to the exponential cut-off assumed for the stellar energetic particle spectrum; this is similar to what \cite{barth_2021} found in their Fig.~4. For a more active/younger star than GJ\,436 it would be reasonable to assume a higher cut-off energy, as motivated by Eq.\,\ref{eq:hillas} which would result in stellar energetic particles dominating up to higher energies \citep[as discussed in][in the context of a young Sun-like star]{rodgers-lee_2021a}. 

Other methods for estimating stellar energetic particle spectra use relationships between solar far-UV \citep{youngblood_2017}, X-ray flare energy \citep{herbst_2019a} or starspot size \citep{herbst_2021} and $>10\,$MeV proton flux. This is similar to our scaling of the overall spectra with stellar wind kinetic energy. The stellar energetic particle spectrum is normalised using these proton fluxes once a spectral shape has been selected. The shape of the spectrum is often based on a solar energetic particle event \citep[e.g. from][]{mewaldt_2005}, as adopted by \citet{rab_2017} and \citet{barth_2021} for instance, or the strongest ground level enhancement event measured on Earth \citep[e.g.][]{herbst_2019,scheucher_2020}. This is the main difference between our model and those based on solar energetic particle events. The spectral shape obtained from our model depends on stellar magnetic field strength which varies from star to star.

As mentioned above, we consider flare accelerated particles rather than those from CMEs as stellar CME properties remain elusive \citep[e.g.][for a recent detection of coronal dimming of X-ray emission due to a CME and references therein]{veronig_2021}. However, \citet{hu_2022} recently modelled the expected stellar energetic particle fluxes from a CME associated with a young solar-like star and found higher maximum particle energies for higher CME speeds. On the other hand, previous stellar wind simulations including CMEs \citep{alvarado-gomez_2018} have suggested that strong solar-like CMEs may be suppressed by a large-scale dipole stellar magnetic field of 75\,G which would affect the corresponding stellar energetic particle fluxes. \citet{fraschetti_2022} showed for the AU Mic system how the stellar energetic particle flux distribution reaching the planetary orbits is strongly altered by the passage of a CME. To complicate matters further, solar CME-CME interactions account for more than 25\% of the major geomagnetic storms observed \citep{zhang_2007}. Using models, \citet{koehn_2022} suggest that successive CMEs can lead to extreme conditions at Earth. Thus, it would be expected that stellar CME-CME interactions will also affect the associated stellar energetic particle fluxes.

\subsection{Exoplanet atmosphere model}
\label{subsec:atmos_profile}
\begin{figure}
	\centering
        \includegraphics[width=0.5\textwidth]{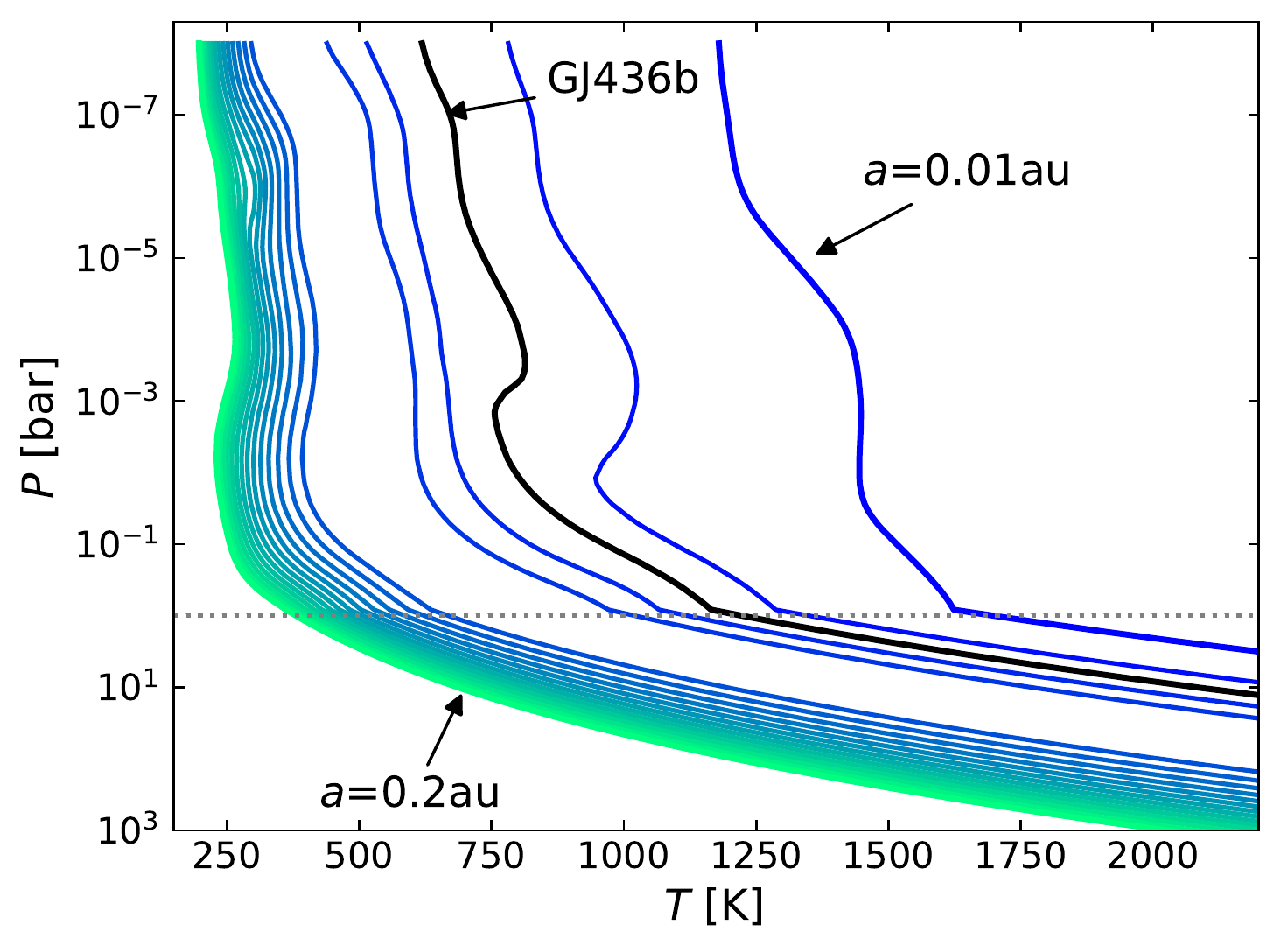}
       	\centering
  \caption{Temperature profiles of a hydrogen-dominated warm Neptune  atmosphere for orbital distances between 0.01\,au (rightmost blue line) and 0.2\,au (leftmost green line) in the GJ\,436 system. The black line represents the temperature profile at the orbital distance of GJ\,436\,b. The profiles were determined using HELIOS when pressure, $P < 1\,$bar, and assuming an adiabat when $P > 1\,$bar, indicated by the grey dotted line.\label{fig:TP-profiles}} 
\end{figure}

The exoplanet atmosphere density ($n(z)$) in units of cm$^{-3}$, is important for the propagation of the energetic particles through the exoplanet atmosphere. The 1D temperature pressure profiles for a Neptune-like planet in the GJ\,436 system are found using the radiative transfer code HELIOS \citep{malik_2017,malik_2019}. We assume an adiabatic atmosphere for pressures, $P>1$\,bar\footnote{We have not included any internal temperature in the HELIOS models. Thus, taking into account that the radiative gradient is proportional to the planet luminosity, no adiabat will occur for our HELIOS models. Setting an internal temperature in HELIOS could not be done for the temperature pressure profiles considered due to non-convergence effects. We instead include an adiabat for $P> 1$\,bar. This is a reasonable approximation given the uncertainties in internal temperatures and opacities for exoplanet atmospheres.}. The density is then calculated using the ideal gas law for the entire atmosphere. The profiles, shown in Fig.\,\ref{fig:TP-profiles}, are calculated for a Neptune-like planet at different orbital distances, $a$, ranging from 0.01-0.2\,au in steps of 0.01\,au which includes GJ\,436\,b's observed orbital distance (i.e. 0.028\,au). 

The required inputs for HELIOS are the planetary parameters (i.e. mass, radius and orbital distance), the opacities and chemical abundances of included species and the stellar spectrum. For the opacities we make use of the line lists from the DACE database \citep{grimm_2015,grimm_2021} and we only take the most relevant species into account (i.e. CH$_4$, CN, CO, CO$_2$, H$_2$O, H$_2$S, NH$_3$, NaH, PH$_3$, SiO, TiO, and VO) considering solar elemental abundance \citep{asplund_2009} in chemical equilibrium  \citep[as described in][Fig.\,3]{louca_2022}. The stellar XUV spectrum is taken from the MUSCLES survey \citep{france_2016,youngblood_2016,loyd_2016,youngblood_2017}. For higher wavelengths PHOENIX models were used and stitched to the XUV spectrum. The stellar and planetary parameters for the GJ\,436 system are given in Table\,\ref{table:stellar_parameters}. For the HELIOS models where the planet had an orbital distance (period) $<$0.069\,au ($<$10 days), the model assumed a tidally locked planet, with no heat-redistribution. Planets with orbital distances $\geq 0.069$\,au were not assumed to be tidally locked anymore and global heat-redistribution was included. The final temperature pressure profiles are averaged over the irradiated hemisphere.

\setlength{\tabcolsep}{2pt}
\begin{table}
\centering
\caption{Stellar and planetary parameters for the GJ\,436 system.}
\begin{tabular}{@{}ccccccc@{}}
\hline

Identifier & $M$ & $R$ & $T_\mathrm{eff}$ & $a$ & $P_{\rm rot}$ & Ref. \\
\hline
 &  & & (K) & (au) & (days) &  \\
\hline
GJ\,436 & $0.452 M_\odot$ & $0.437 R_\odot$ & 3479 & -& - & 1\\
\hline
GJ\,436\,b & $0.08 M_\mathrm{J}$	& $0.372 R_\mathrm{J}$	& 879	& 0.028 & 2.64 & 2\\
\hline 
\label{table:stellar_parameters}
\end{tabular}

(1) \citet{knutson_2011}; (2) \citet{butler_2004}. The effective temperature of GJ 436 b, as shown in this table, is derived using radiative transfer calculations with HELIOS. 
\end{table}

\subsection{Energetic particle transport in exoplanet atmospheres}
\label{subsec:cr_atmos}
The energetic particle propagation through the exoplanet atmosphere is performed using a Monte Carlo model \citep[from][]{rimmer_2012,rimmer_2013}. Here, we briefly outline the method which is described in detail in \citet{rimmer_2012,rimmer_2013}. The model is based on the continuous slowing down approximation where the energetic particles travel through a column density, $n(z)dz$, of the exoplanet atmosphere. Here, we take $5\times 10^5$ particles to ensure that the fluctuations per energy bin are negligible. In order to calculate the energetic particle energy losses for each height element, $dz$, in the exoplanet atmosphere, each energetic particle is assigned two values: an individual kinetic energy, $E$, such that the energetic particle energies are normalised to match the top-of-atmosphere fluxes shown in Fig.\,\ref{fig:cr_fluxes} and a random number, $N$, of uniform distribution between 0 and 1. The random number is compared with the energetic particle `optical depth', $\sigma^{\rm ion}_{\rm p,X} n(z)dz$, for the cell $dz$ where $\sigma^{\rm ion}_{\rm p,X}(E)$ is the ionisation cross section. The subscript `$\rm X$' refers to the ionisation cross sections for $\mathrm{H_2}$ and He from \citet{rudd_1985} and \citet{padovani_2009}. If $N<\sigma^{\rm ion}_{\rm p,X} n(z)dz$, the energetic particle collides and loses energy. This requires that $\sigma_{\rm max,X} n(z)dz<1$, where $\sigma_{\rm max,X}$ is the maximum cross section value for $\mathrm{H_2}$ and He. Following \citet{rimmer_2012}, based on \citet{cravens_1978}, the average energy loss per collision, $\overline{W}$, is given by $\overline{W} = 7.92E^{0.082}+4.76$, where $\overline{W}$ and $E$ have units of eV. For this model, we assume that the atmosphere is composed of 80\% H$_2$ and 20\% He. We have adopted the same expression for $\overline{W}$ for H$_2$ and He. This step is repeated for each height element and new random numbers are assigned. 

The transport for stellar energetic particles and Galactic cosmic rays is treated separately. This is because if the fluxes are combined, the energetic particle flux is dominated by the stellar energetic particles and the high-energy Galactic cosmic rays are poorly sampled by the Monte Carlo method. 

In our model we assume that the energetic particle fluxes are isotropically distributed at the top of the atmosphere, as shown in Fig.\,\ref{fig:sketch}. We also do not account for the effect of a planetary magnetic field. First, a planetary magnetic field would act to deflect low-energy energetic particles towards the magnetic poles. Thus, the energetic particle fluxes would no longer be isotropically distributed at the top of the atmosphere and only particles with Larmor radii comparable to or larger than the planetary radius would penetrate the magnetosphere. Second, while diffusion tends to isotropise energetic particle fluxes, such as for Galactic cosmic rays, the stellar energetic particles are more influenced by advection processes at such small orbital distances \citep[see Fig.\,4 from][]{mesquita_2021}. Thus, for close-in exoplanets the stellar energetic particle fluxes may be higher on the side of the exoplanet facing the star which we do not account for in our model.

\begin{figure}
	\centering
        \includegraphics[width=0.5\textwidth]{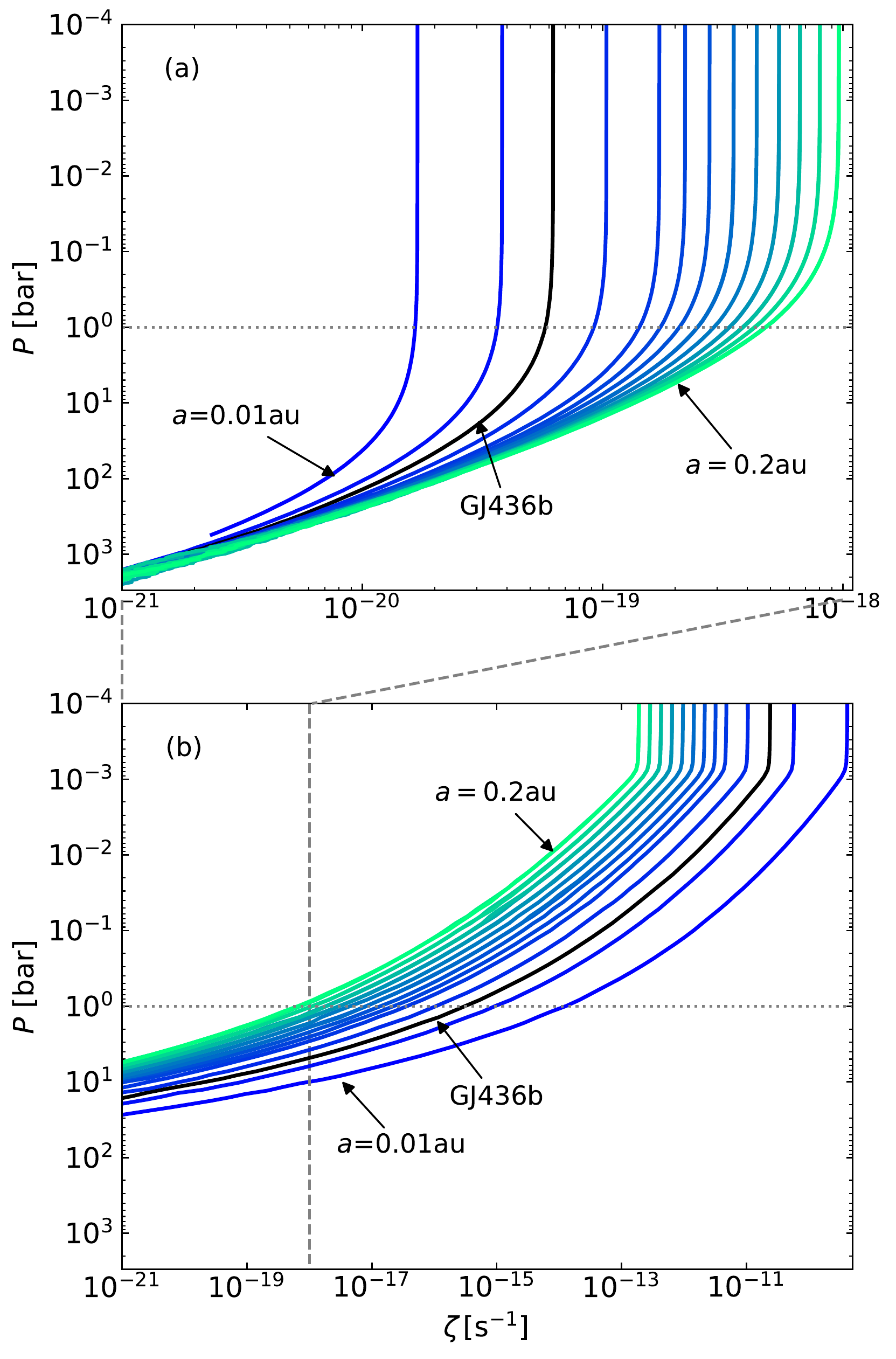}
       	\centering
  \caption{The $\mathrm{H_2}$ ionisation rates are shown as a function of pressure in an exoplanet atmosphere at different orbital distances in the GJ\,436 system for (a) Galactic cosmic rays and (b) stellar energetic particles. The grey dotted line denotes $P=1\,$bar. It is important to note that the $x$-axis ranges are different between the two plots which is indicated by the grey dashed lines between the panels. \label{fig:zeta_cr}} 
\end{figure}

The quantities, presented in Section\,\ref{sec:results}, relevant for chemical modelling of exoplanet atmospheres are the ionisation rate, $\zeta\,$($\rm s^{-1}$), and the ion-pair production rate, $Q = n\zeta\,(\rm cm^{-3}\,s^{-1})$. Following \citet{padovani_2009}, the ionisation rate of molecular hydrogen, $\zeta_{\rm H_2}$, by protons is calculated from the energetic particle differential intensities, $j(E)$, as:
\begin{equation}
\zeta_{\rm H_2} = 4\pi\int^{E_{\rm max}}_{I(\rm H_2)} j(E)[1+\phi(E)]\sigma^{\rm ion}_{\rm p,H_2}(E)\,dE 
\end{equation}
where $I(\mathrm{H}_2)=15.603\,$eV is the ionisation potential of $\rm H_2$ and $E_{\rm max}$ is the maximum energy of the energetic particle spectrum. The parameter, $\phi(E)$, is a correction factor that accounts for additional ionisation of $\rm H_2$ by secondary electrons given by:
\begin{equation}
\phi(E) = \frac{1}{\sigma^{\rm ion}_{\rm p,H_2}(E)}\int^{E_{\rm e,max}}_{I(\rm H_2)} \underbrace{\frac{1}{\sigma^{\rm ion}_{\rm p,H_2}(E)}\frac{d\sigma^{\rm ion}_{\rm p,H_2}(E)}{dE_{\rm e}}}_{P_{\rm e}(E,E_{\rm e})} \sigma^{\rm ion}_{\rm e}(E_{\rm e})\,dE_{\rm e},
\end{equation}
where $E_{\rm e}$ (eV) is the secondary electron energy, $\sigma^{\rm ion}_{\rm e}(E_{\rm e})$ (cm$^2$) is the ionisation cross section of ${\rm H_2}$ by electrons and $P_{\rm e}(E,E_{\rm e})$ is the probability density that a secondary electron with energy $E_{\rm e}$ is produced as a result of an initial ionising proton of energy $E$. This probability density is represented in terms of the total ionisation cross-section ($\sigma^{\rm ion}_{\rm p,H_2}(E)$, cm$^2$), and differential cross-section ($d\sigma^{\rm ion}_{\rm p,H_2}(E)/dE_{\rm e}$, cm$^2$ MeV$^{-1}$). From \citet{cravens_1975},
\begin{equation}
\frac{d\sigma^{\rm ion}_{\rm p,H_2}(E)}{dE_{\rm e}} = \frac{A(E)}{1+\left(\frac{E_{\rm e}}{E_0} \right)^{2.1}},
\end{equation}
where $E_0=8.3\,$eV. Given that, by definition:
\begin{equation}
\displaystyle\int\limits_0^\infty P_{\rm e}(E,E_{\rm e})\,dE_{\rm e} = 1,\notag
\end{equation} 
$A(E)$ can be expressed as $A(E) = C\sigma^{\rm ion}_{\rm p,H_2}(E)$ where the value of $C$ is determined from numerical integration.

In addition to being important for chemical modelling, energetic particles will impact life on other planets leading to increased mutation rates \citep[for instance see discussion in][]{scalo_2007,dartnell_2011}. The equivalent dose rate, $\dot{D}$\,(Sv\,s$^{-1}$), which is the energy absorbed per unit time and mass, provides a measure of how damaging energetic particle fluxes are for life-forms. We estimate the skin-depth equivalent dose rate from primary protons as:
\begin{equation}
\dot{D} \sim 2\pi W \int^{E_{\rm max}}_{E_{\rm min}} \frac{E\,j(E)}{R(E)}\,dE
\label{eq:dose}
\end{equation}  
where $W=2$ is the radiation weighting factor for protons and $R(E)\,({\rm g\,cm^{-2}})$ is the proton range in water\footnote{https://physics.nist.gov/PhysRefData/Star/Text/PSTAR.html.}. The quantity $E\,j(E)$ in Eq.\,\ref{eq:dose} represents the particle number flux in an energy bin, $dE$. We have assumed that the energetic particles only impact from above. In Section\,\ref{subsec:d}, we present $\dot{D}$ in units of ${\rm mSv\,day^{-1}}$.

\section{Results}
\label{sec:results}
Here we present our results for the energetic particle propagation in the atmosphere of a GJ\,436\,b-like planet at different orbital distances. Using the energetic particle fluxes (shown in Fig.\,\ref{fig:cr_fluxes}) and the atmospheric profiles (Fig.\,\ref{fig:TP-profiles}), we calculate the ionisation rate (Section\,\ref{subsec:zeta}), the ion-pair production rate (Section\,\ref{subsec:q}) and the skin-depth equivalent dose rate (Section\,\ref{subsec:d}).

\begin{figure*}%
	\centering
    \subfigure[]{%
        \includegraphics[width=0.5\textwidth]{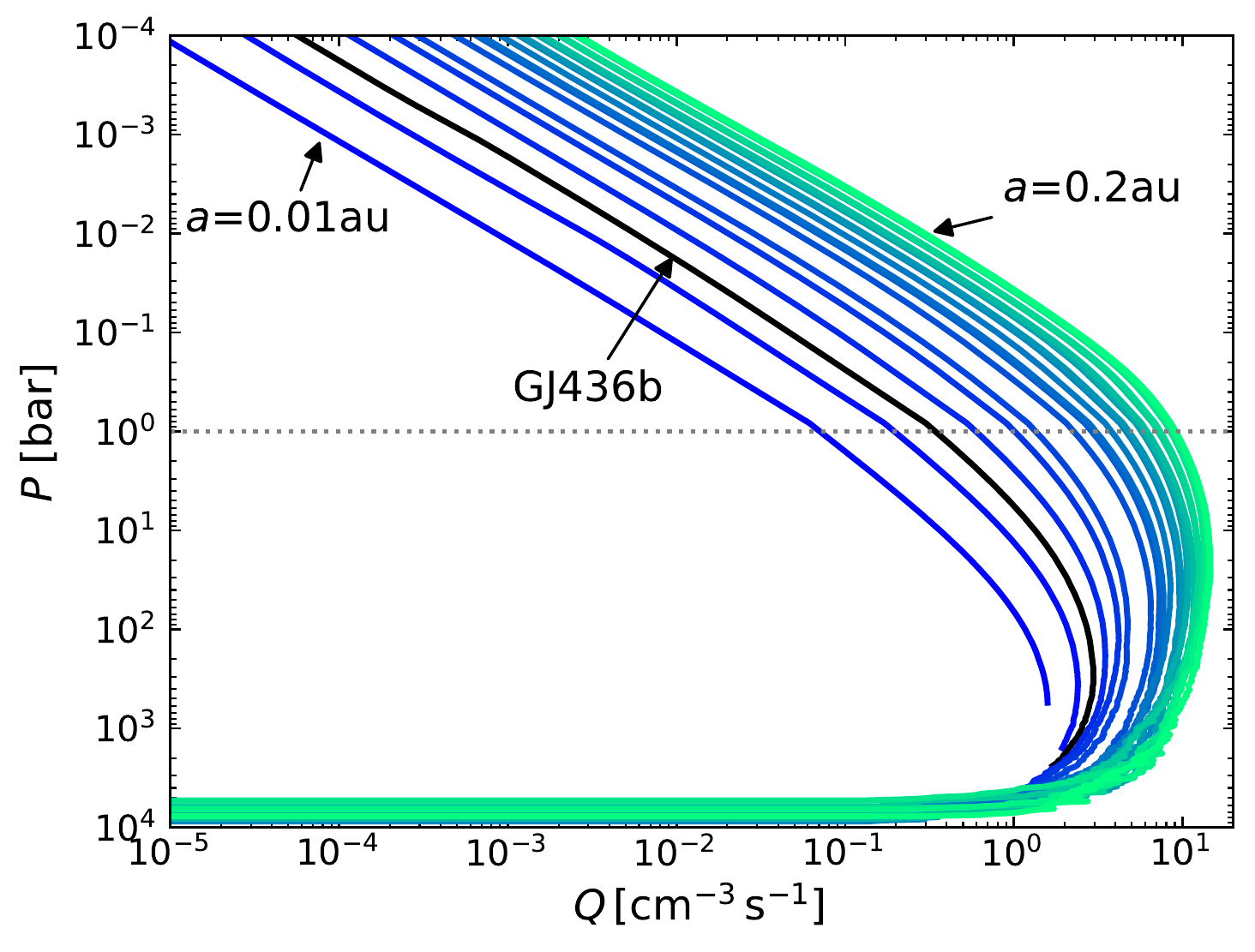}
       	\centering
\label{fig:q_gcr}}%
~
	\centering
    \subfigure[]{%
        \includegraphics[width=0.5\textwidth]{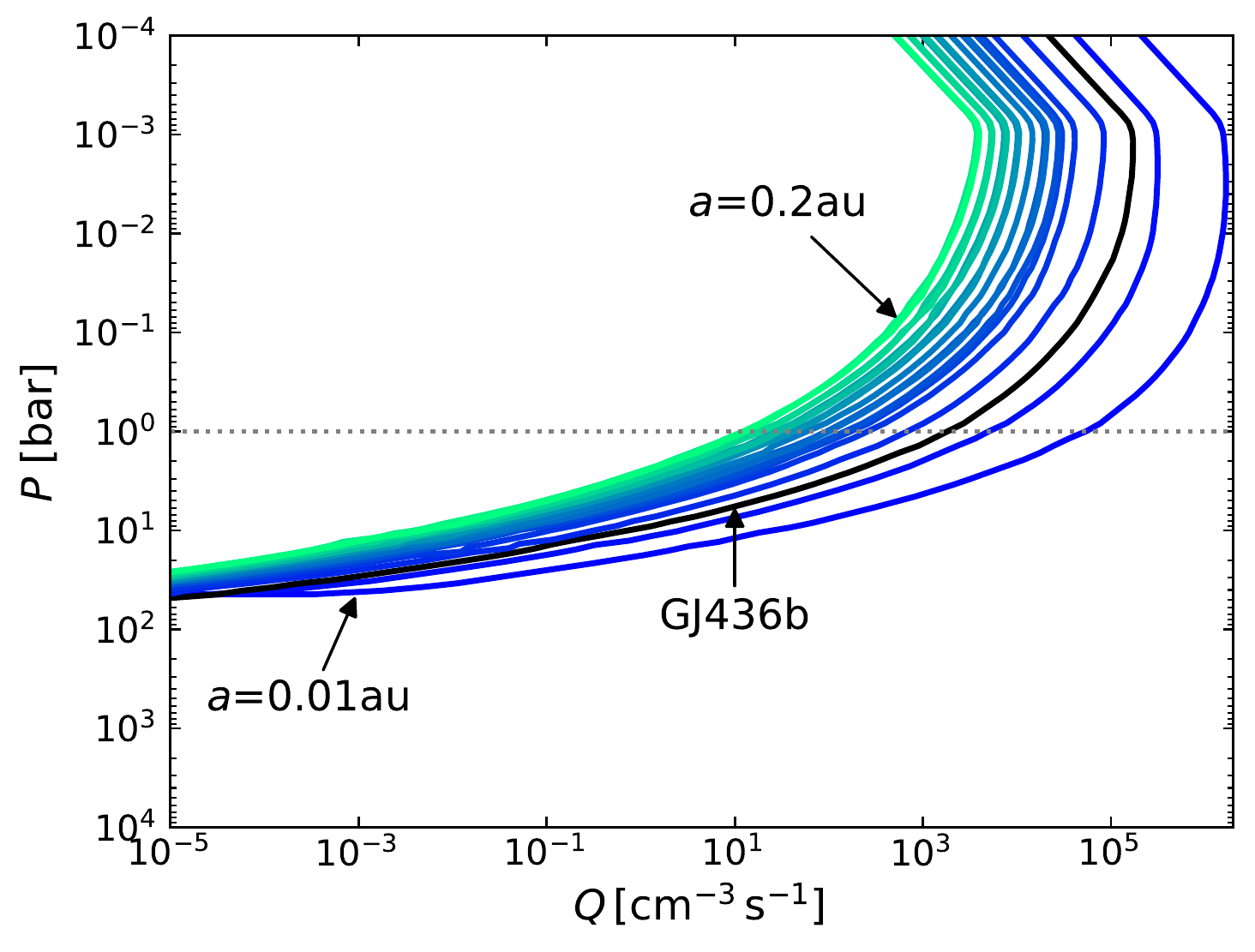}
       	\centering
\label{fig:q_stcr}}%
    \caption{The ion-pair production rate, $Q$, as a function of pressure in an exoplanet atmosphere for different orbital distances in the GJ\,436 system for (a) Galactic cosmic rays and (b) stellar energetic particles. The grey dotted line denotes $P=1\,$bar. The linestyles are the same as for Fig.\,\ref{fig:zeta_cr}.} 
    \label{fig:q_cr}%
\end{figure*}

\subsection{Ionisation rates}
\label{subsec:zeta}
Fig.\,\ref{fig:zeta_cr} shows $\zeta_{\rm H_2}$ from Galactic cosmic rays (Fig.\,\ref{fig:zeta_cr}(a)) and stellar energetic particles (Fig.\,\ref{fig:zeta_cr}(b)) as a function of pressure for different orbital distances ($a=0.01 - 0.2$\,au) with different coloured lines. The black line represents $\zeta_{\rm H_2}$ at the orbital distance of GJ\,436\,b. The grey dotted line denotes $P=1$\,bar. At the top of the atmosphere, $\zeta_{\rm H_2}$ from Galactic cosmic rays varies by approximately two orders of magnitude ranging from $\sim 1.7 \times 10^{-20} - 10^{-18}\,{\rm s^{-1}}$ between $a=0.01-0.2\,$au, shown in Fig.\,\ref{fig:zeta_cr}(a). In comparison, the LIS for Galactic cosmic ray protons results in a higher value of $\zeta_{\rm H_2}\sim 1.6 \times 10^{-17}\,{\rm s^{-1}}$ \citep[estimated here as $2\zeta_{\rm H}$ calculated in][from \textit{Voyager} data]{cummings_2016}. For the orbital distances considered in Fig.\,\ref{fig:zeta_cr}(a), the top-of-atmosphere $\zeta_{\rm H_2}$ from Galactic cosmic rays is largest at the largest orbital distance of 0.2\,au. This is because the Galactic cosmic ray fluxes are highest at the LIS values in the ISM which are then suppressed as they travel inwards through the stellar wind. The Galactic cosmic rays diffuse into the stellar system from the ISM. At the same time, the Galactic cosmic ray fluxes are suppressed by advection processes due to the expanding magnetised stellar wind. In Fig.\,\ref{fig:zeta_cr}(a), $\zeta_{\rm H_2}$ starts to decrease rapidly between $P\sim 10^{-1}-10^1$\,bar, depending on the orbital distance considered. This illustrates that the Galactic cosmic rays travel essentially unimpeded through the exoplanet atmosphere in regions where the density is relatively low ($P\leq10^{-1}\,$bar). 

For stellar energetic particles (Fig.\,\ref{fig:zeta_cr}(b)), $\zeta_{\rm H_2}$ begins to decrease rapidly at $P\sim 10^{-3}$\,bar. The top-of-atmosphere $\zeta_{\rm H_2}$ values vary by approximately three orders of magnitude ranging from $\sim 2 \times 10^{-13} - 4\times 10^{-10}\,{\rm s^{-1}}$. In contrast to $\zeta_{\rm H_2}$ from Galactic cosmic rays, the highest top-of-atmosphere $\zeta_{\rm H_2}$ values for the stellar energetic particles are found at the smallest orbital distances. This is because the stellar energetic particles originate from a $\sim$point source close to the stellar surface, causing the fluxes to decrease with increasing orbital distance (Fig.\,\ref{fig:stcr_fluxes}) as they are advected by and diffuse through the stellar wind in our model. As mentioned above, the reverse is true for Galactic cosmic rays which have the lowest fluxes closest to the star (Fig.\,\ref{fig:gcr_fluxes}) as they originate from the ISM. The stellar energetic particle values for $\zeta_{\rm H_2}$ decrease much more rapidly with increasing pressure than the Galactic cosmic ray values (compare the black lines from Figs.\,\ref{fig:zeta_cr}(a) and \ref{fig:zeta_cr}(b), for instance). This is because the stellar energetic particles are predominantly at $\sim$MeV energies which have a higher energy loss rate than $\gtrsim$GeV energy cosmic rays. Thus, in the upper atmosphere stellar energetic particles dominate while in deeper parts ($P\gtrsim 100\,$bar) Galactic cosmic rays are more important. Transmission spectroscopy with JWST and Ariel will likely probe regions between $P\sim 10^{-4}-10^{-1}$\,bar for $\rm{H_2}$-rich atmospheres \citep{welbanks_2019}. 

Thus, for the atmospheres and orbital distances ($a=0.01-0.2\,$au) that we consider here for the GJ\,436 system, transmission spectroscopy with JWST and Ariel would probe regions where stellar energetic particles are more likely to impact the chemistry than Galactic cosmic rays (also see Section\,\ref{subsec:q} where the ion-pair production rates are discussed). While ionisation and dissociation by energetic particles, including photons, are likely the main sources of chemical disequilibrium in the upper atmosphere, vertical mixing and chemical quenching are likely the dominant sources of disequilibrium chemistry deeper in the atmosphere \citep{tsai_2018, ohno_2022}. This trend will be dependent on the specific nature of the temperature profile. Cooler atmospheres, especially those of rocky planets, may have other sources of disequilibrium \citep[e.g. degassing, deposition, condensation, escape,][]{ noack_2014,tosi_2017,ortenzi_2020,zahnle_2020}. Very hot atmospheres may never experience chemical quenching \citep{kitzmann_2018}.

The total $\zeta_{\rm H_2}$ from energetic particles as a function of pressure can be calculated by summing the individual $\zeta_{\rm H_2}$ values shown in Fig.\,\ref{fig:zeta_cr} for Galactic cosmic rays and stellar energetic particles\footnote{Alternatively, the top-of-atmosphere energetic particle fluxes could have been summed. However, as described in Section\,\ref{subsec:cr_atmos}, we treat the stellar energetic particle and Galactic cosmic ray transport separately.}. These ionisation rates can be used to study the energetic particle-induced chemistry in an exoplanet atmosphere, similar to \citet{barth_2021}. \citet{barth_2021} studied the hot Jupiter, HD\,189733\,b, orbiting its K dwarf host star at 0.031\,au, similar to GJ\,436\,b's orbital distance of 0.028\,au. As discussed in Section\,\ref{subsec:step_fluxes}, the fact that the adopted stellar energetic particle spectra are different, in combination with the different temperature-pressure profiles, may result in different chemical tracers for energetic particles. The main difference in the stellar energetic particle spectra is that our spectra has more $\sim$GeV energy particles at 0.03\,au, which penetrate further into an exoplanet atmosphere.

As shown in Fig.\,\ref{fig:zeta_cr}, the top-of-atmosphere values $\zeta_{\rm H_2}$ values from stellar energetic particles are all larger than those due to Galactic cosmic rays for $a=0.01-0.2\,$au. However, at 10\,au we find that the value of $\zeta_{\rm H_2}=7\times 10^{-18}\,\mathrm{s^{-1}}$ due to stellar energetic particles is the same as that obtained from Galactic cosmic rays. The Galactic cosmic ray fluxes, and the resulting ionisation rates, are highest at large orbital distances where they have travelled through relatively little of the stellar wind. Galactic cosmic rays from the ISM lose energy as they diffuse through the stellar wind. At the same time, stellar energetic particle fluxes decrease with distance from the star. 

Thus, for an exoplanet at an orbital distance $>10$\,au in the GJ\,436 system it is more likely that any observable chemical signatures of energetic particles from the upper atmosphere will be due to Galactic cosmic rays rather than stellar energetic particles. Additionally, as mentioned above, the (higher energy) Galactic cosmic rays lose their energy less rapidly with increasing atmospheric pressure/density in comparison to the (lower energy) stellar energetic particles. This means, for an exoplanet at an orbital distance $>10$\,au in the GJ\,436 system, that Galactic cosmic rays will likely be the dominant source of ionisation due to energetic particles at all heights in the atmosphere. This should be true for many systems, as discussed in \citet{rodgers-lee_2020b}, though the exact orbital distance where this occurs will be system-dependent due to different stellar wind conditions. For instance, Fig.3 of \citet{mesquita_2021} shows the effect of assuming two different stellar wind models on the Galactic cosmic ray spectra at various orbital distances for the GJ\,436 system. The difference between these stellar wind models is that for one model the wind is more magnetically-dominated (Case A adopted here), whereas the other is more thermally-dominated (Case B). Thus, the value of $\zeta_{\rm H_2}$ from Galactic cosmic rays at 10\,au would decrease somewhat if we had adopted the thermally-dominated stellar wind, largely due to the larger astrosphere resulting from this stellar wind model. For other systems which are dominated more by advective processes \citep[see][for instance]{rodgers-lee_2021b}, the difference in Galactic cosmic ray $\zeta_{\rm H_2}$ values at large distances assuming different astrosphere sizes would be more significant.

\subsection{Ion-pair production rates}
\label{subsec:q}
The ion-pair production rate, $Q=n\zeta$, convolves the number density of the exoplanet atmosphere with the ionisation rate. It gives an indication of where in the exoplanet atmosphere the energetic particles will be important for chemistry by creating the most ions. Thus, high $Q$ values correspond to regions where the energetic particles should be most important. Fig.\,\ref{fig:q_gcr} and \ref{fig:q_stcr} show the ion-pair production rate, $Q$, as a function of pressure from Galactic cosmic rays and stellar energetic particles, respectively for different orbital distances. The maximum Galactic cosmic ray $Q$ value occurs at $P\sim 10-10^2\,$bar (rightmost green line in Fig.\,\ref{fig:q_gcr}). For stellar energetic particles instead, the maximum $Q$ value occurs much higher in the atmosphere between  
$P\sim 10^{-3}-10^{-2}\,$bar, with a much greater value (rightmost blue line). Here, we have ignored the diffusion of ions. 

\begin{figure}
	\centering
        \includegraphics[width=0.5\textwidth]{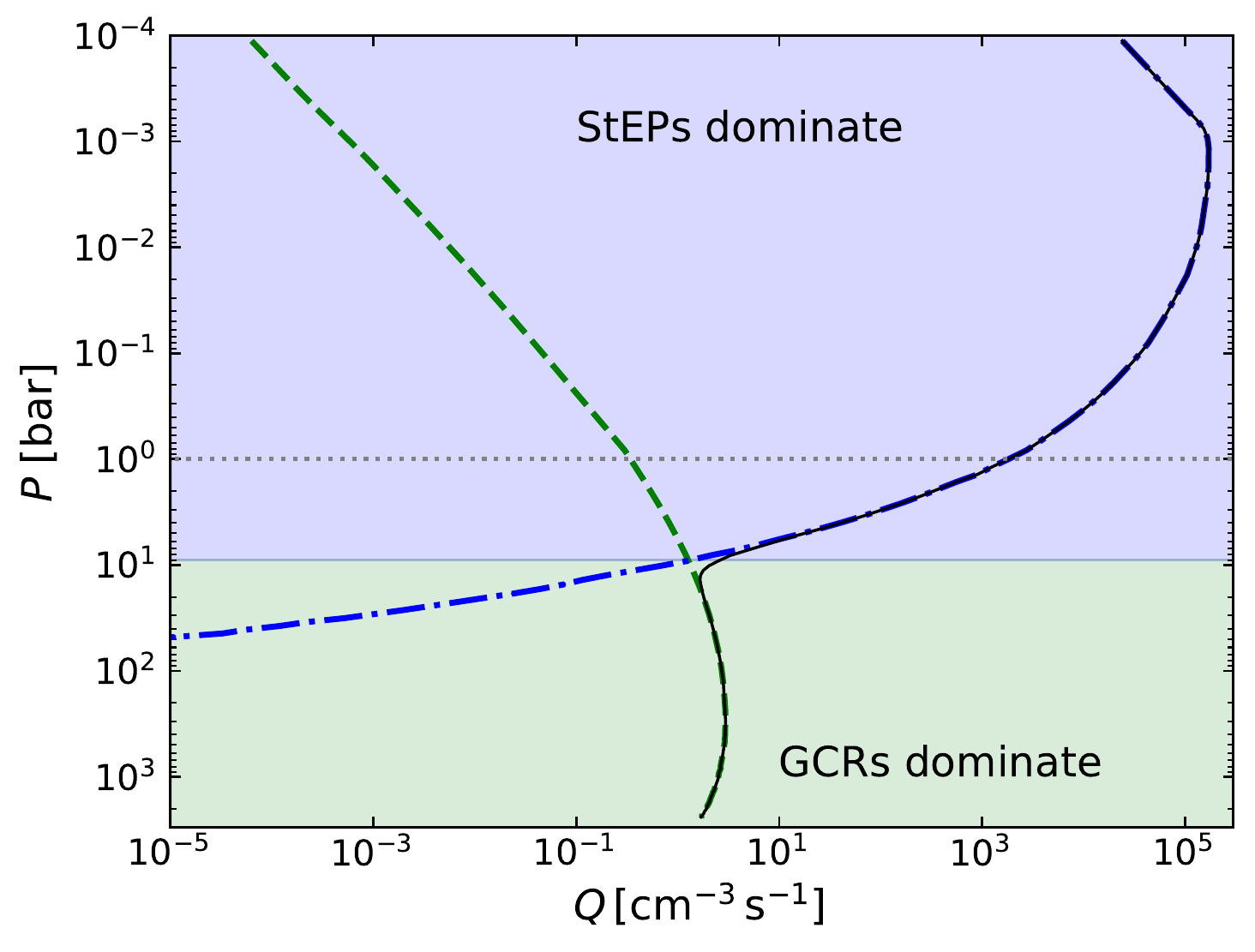}
       	\centering
  \caption{The ion-pair production rate, $Q$,  as a function of pressure at the orbital distance of GJ\,436b for Galactic cosmic rays (labelled as `GCRs', green dashed line) and stellar energetic particles (labelled as `StEPs', blue dot-dashed line). The solid black line represents the total $Q$ value. The grey dotted line denotes $P=1\,$bar. \label{fig:q_gj436b}} 
\end{figure}

For comparison, Fig.\,\ref{fig:q_gj436b} shows the ion-pair production rate for Galactic cosmic rays ($Q^{\rm GCR}$, green dashed line) and stellar energetic particles ($Q^{\rm StEP}$, blue dot-dashed line) at the orbital distance of GJ\,436\,b as a function of pressure. The thin solid black line represents the combination of both. The blue shaded region illustrates where $Q^{\rm StEP}>Q^{\rm GCR}$ and the green shaded region shows where $Q^{\rm GCR}>Q^{\rm StEP}$ which occurs at $\sim$1\,bar for GJ\,436\,b. The fact that Galactic cosmic rays are more important than stellar energetic particles at $10-10^3$\,bar is interesting because it has been suggested that a $\sim10^2\,$bar atmosphere is expected for a post-impact early Earth atmosphere \citep{zahnle_2020,itcovitz_2022}. This suggests that Galactic cosmic rays could have been the dominant ionisation source deep in the atmosphere at this time.

\citet{herbst_2019} and \citet{scheucher_2020} calculate the ion-pair production rate from Galactic cosmic rays and solar/stellar energetic particles for an Earth-like atmosphere and for Prox Cen\,b, respectively. Fig.\,6 of \citet{herbst_2019} shows broadly the same behaviour as our results: $Q^{\rm GCR}$ peaks at a lower  value than $Q^{\rm StEP}$, but occurs at a higher pressure/lower altitude. The main difference is that for our model atmosphere for GJ\,436\,b, Galactic cosmic rays dominate for a broad range of high pressures. The fact that our maximum $Q^{\rm StEP}$ value peaks at $\sim10^{-3}\,$bar reflects that our assumed stellar energetic particle spectrum decays exponentially at a relatively low cosmic ray kinetic energy. For a more active star, we would expect the maximum $Q^{\rm StEP}$ value to occur at a higher pressure. 

In terms of the range of values for $Q^{\rm StEP}$ and $Q^{\rm GCR}$, our peak $Q^{\rm GCR}$ value is lower than that found in \citet{herbst_2019} which is reasonable since our top-of-atmosphere Galactic cosmic ray spectrum (Fig.\,\ref{fig:gcr_fluxes}) is reduced in comparison to values observed at Earth. Our peak $Q^{\rm StEP}$ value lies between the maximum values found in \citet{herbst_2019} and \citet{scheucher_2020} which is again consistent since our stellar energetic particle fluxes at $\sim 10-100\,$MeV energies lie between those assumed for the Earth-like atmosphere and for Prox Cen\,b. Our stellar energetic particle fluxes are higher for GJ\,436\,b mainly because it orbits much closer to its star than the Earth orbits the Sun, while they are lower than for Prox Cen\,b. At the same time, it is complex to compare values since $Q$ depends not only on the incident energetic particle fluxes but also on the density profile of the planetary atmosphere. Additionally, \citet{herbst_2019} include the effect of an Earth-like planetary magnetic field which introduces a low-energy particle cut-off, i.e. energetic particles below a certain energy (510\,MeV in their case for latitudes of around 60$^\circ$) cannot penetrate the planet's magnetosphere and do not contribute to $Q$. In this context, this indicates that our results represent the largest possible values for $Q$ and that exoplanetary magnetic fields would generally result in a decrease of the values obtained for $Q$. It is likely that the $Q^{\rm StEP}$ values would be more affected than $Q^{\rm GCR}$. This is because the top-of-atmosphere stellar energetic particle spectra are dominated by low-energy particles, whereas the peaks of the Galactic cosmic ray spectra occur closer to GeV energies (see Fig.\,\ref{fig:cr_fluxes}).

\subsection{Skin-depth equivalent dose rates}
\label{subsec:d}
\begin{figure}
	\centering
        \includegraphics[width=0.5\textwidth]{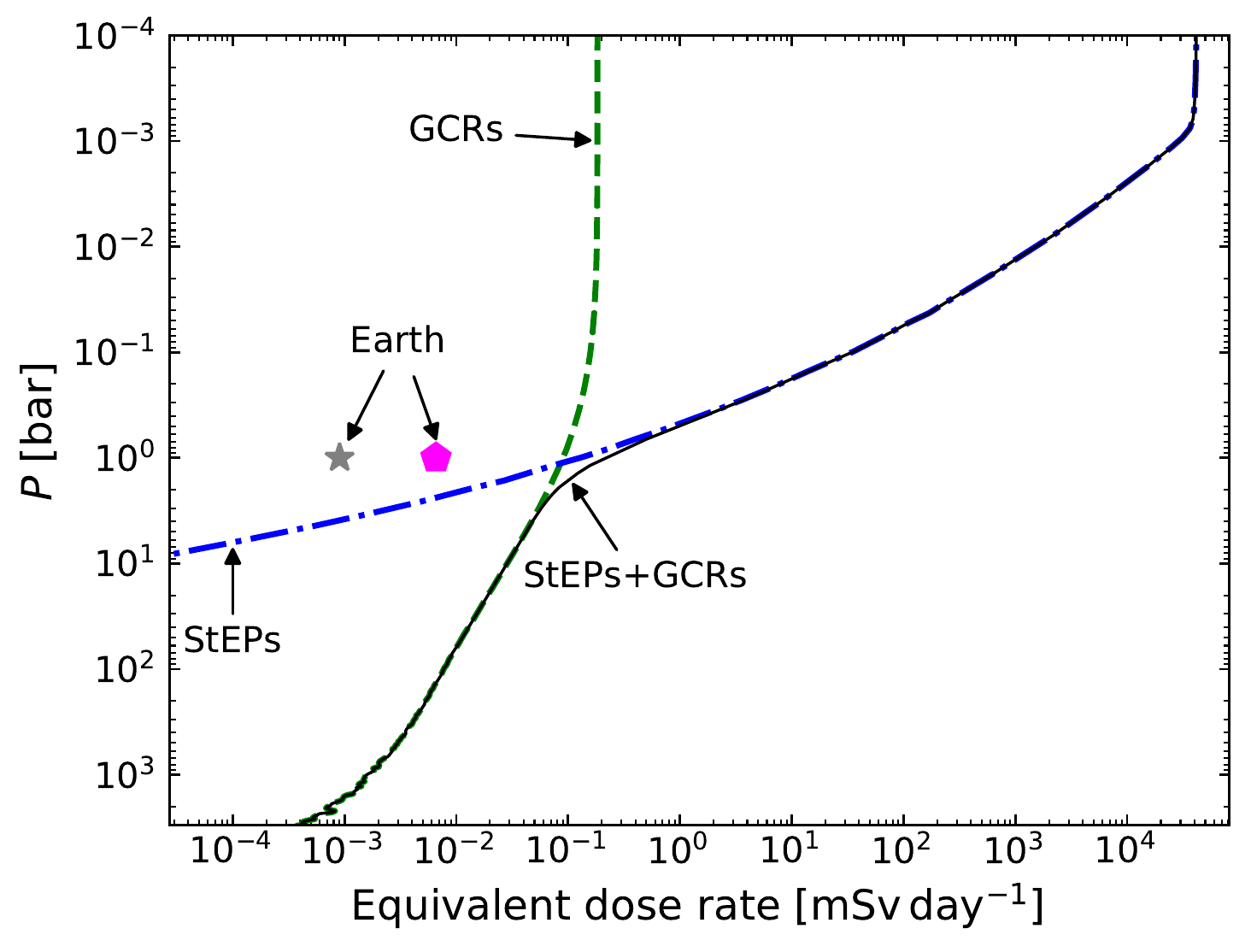}
       	\centering
  \caption{The skin-depth equivalent dose rate ($\dot D$) is plotted as a function of pressure for a warm Neptune exoplanet atmosphere at an orbital distance of 0.2\,au, the inner edge of the habitable zone for the GJ\,436 system. The green dashed line represents $\dot D$ for Galactic cosmic rays (labelled as `GCRs') and the blue dot-dashed line represents $\dot D$ for stellar energetic particles (labelled as `StEPs').The solid black line represents the total skin-depth equivalent dose rate. The grey star and the magenta pentagon represent the equivalent dose rate at Earth's surface due to Galactic cosmic rays and background radiation sources, respectively.\label{fig:dose}} 
\end{figure}

While warm Neptune exoplanets are not expected to be habitable we can calculate the skin-depth equivalent dose rate for these planets. In the future, the same procedure can be applied to stars like GJ\,436 that could have terrestrial planets in their habitable zones. Fig.\,\ref{fig:dose} shows the skin-depth equivalent dose rate from energetic particles as a function of pressure for Galactic cosmic rays (green dashed line) and stellar energetic particles (blue dot dashed line) at $a=0.2\,$au, which is the inner edge of the habitable zone for the GJ\,436 system. The thin black line represents the combination of both. The grey star and magenta pentagon represent the average equivalent dose rate at Earth's surface due to Galactic cosmic rays ($9\times 10^{-4}$mSv$\,\rm day^{-1}$, including however secondary electrons, muons and $\gamma-$rays) and background radiation sources ($7\times 10^{-3}$mSv$\,\rm day^{-1}$). Here we have used the temperature pressure profile at 0.2\,au, described in Section\,\ref{subsec:atmos_profile}, assuming the same planetary mass and radius as before.

For Galactic cosmic rays, the equivalent dose rate is approximately constant for $P<1\,$bar. Again in comparison to \citet{herbst_2019}, our maximum equivalent dose rates are lower (higher) for Galactic cosmic rays (stellar energetic particles). It is important to note that \citet{herbst_2019} consider a terrestrial planet atmosphere whereas we are still considering a hydrogen dominated atmosphere which affects the energetic particle transport in the exoplanet atmosphere. In turn, this will impact the dose rates that are calculated. 

For these equivalent dose rates, and for chemical modelling that include our energetic particle ionisation rates, it is particularly important to consider how constant these rates would be in time. We discuss this further in Section\,\ref{subsec:continuous}. While indirect detections of exoplanetary magnetic fields remain challenging, Fig.\,9 of \citet{griessmeier_2016} shows that the biological dose rate decreased significantly with increasing magnetic moments for two different atmospheric column depths. This indicates that our results represent an upper limit in the case that the exoplanet possesses a magnetic field.

\begin{figure}
	\centering
        \includegraphics[width=0.5\textwidth]{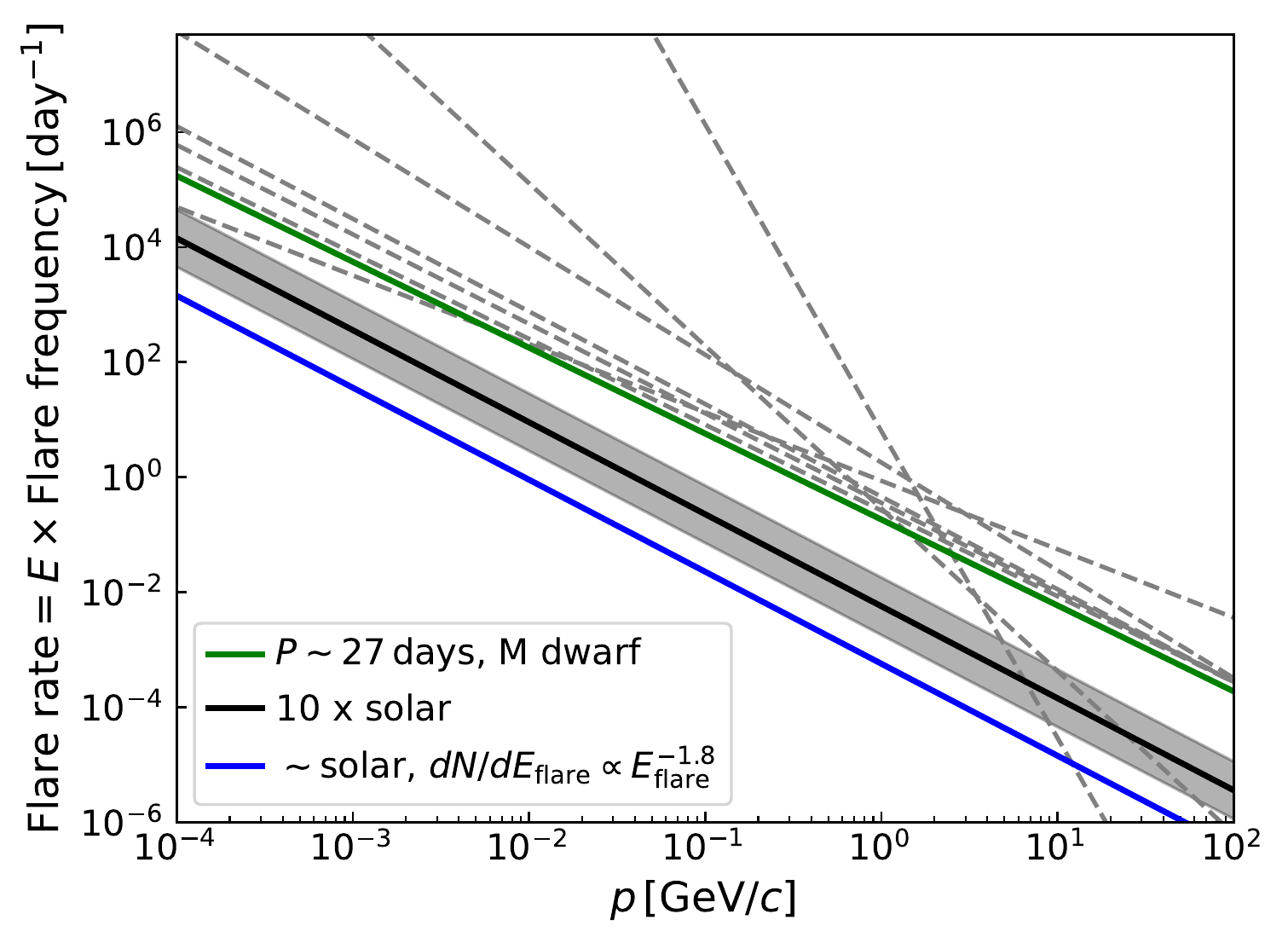}
       	\centering
  \caption{ Flare rates are plotted as a function of the maximum momentum of the stellar energetic particles accelerated. The solid blue line represents a fit to the solar flare frequency \citep[from][]{maehara_2015}. The solid black line is 10 times the solar flare rate and the grey shaded region is $\pm0.5\,$dex of this value. The solid green line is a flare rate from \citet{gunther_2020} for an M dwarf with a rotation period of $\sim27\,$days. The grey dashed lines are flare rates for M dwarfs from \citet{gunther_2020} without a measured rotation period with a spectral type similar to GJ\,436.\label{fig:pmax}} 
\end{figure}

\section{Discussion}
\label{sec:discussion}
\subsection{How continuous are stellar energetic particle fluxes?}
\label{subsec:continuous}
Stellar energetic particle fluxes may vary on timescales of days, depending on stellar activity. Galactic cosmic ray fluxes are more likely to be approximately constant on these timescales. Instead, they vary with the solar cycle \citep[e.g. Fig.3 from][]{vos_2015}, for instance, and will also vary on stellar evolutionary timescales \citep{rodgers-lee_2020b}. Similar to \citet{rodgers-lee_2021a}, we plot the flare rate as a function of the maximum momentum of the stellar energetic particles in Fig.\,\ref{fig:pmax}. This is based on the relation between flare frequency and flare energy ($dN/dE_{\rm flare} \propto E_{\rm flare}^b$, where the power law index $b$ varies from star to star). Flare energies are related to the stellar energetic particle maximum momenta via $p_{\rm max}\sim E_{\rm flare}^{1/2}$, normalised for $\sim$GeV energetic particle energies at $E_{\rm flare} \sim 10^{33}\,$erg and assuming that $E_{\rm flare}\sim B_\star^2$ \citep[see][for more details]{rodgers-lee_2021a}. The underlying relation between flare rate and energy is plotted in Fig.\,\ref{fig:flare}.

The relation between flare rate and energy, and therefore the relation between flare rate and maximum stellar energetic particle momentum is unknown for GJ\,436. Thus, in Fig.\,\ref{fig:pmax}, we plot the flare rate versus maximum stellar energetic particle momentum for the Sun and a number of M dwarfs. In Fig.\,\ref{fig:pmax}, the solid blue line is a fit to the solar flare frequency versus flare energy for a broad range of flare energies using solar observations and Kepler observations of superflares from solar-type stars \citep{maehara_2015}. The solid black line is 10 times the solar flare frequency fit (i.e. $10\times E\,dN/dE_{\rm flare}$, where $dN/dE_{\rm flare} \propto E_{\rm flare}^{-1.8}$) from \citet{maehara_2015}, which is what would be expected for a solar-type star with a rotation period between $5-10$\,days. The grey shaded region is $\pm 0.5\,$dex of this value. The solid green line reflects a fit of the flare rate versus energy for an M dwarf with a rotation period of $\sim$27\,days from \citet{gunther_2020} using TESS data. This M dwarf is selected because it is relatively similar to GJ\,436 in terms of stellar effective temperature (3850\,K) and has a relatively long rotation period. The grey dashed lines are a number of flare rate fits for M dwarfs with similar spectral types to GJ\,436 but with unknown rotation periods \citep[also from][]{gunther_2020}. 

Using 10 times the solar flare rate, Fig.\,\ref{fig:pmax} shows that the value of $p_{\rm max}$=0.3\,GeV$/c$ (kinetic energy of 50\,MeV) assumed for GJ\,436 corresponds to a flare rate of 0.04\,day$^{-1}$, i.e. a flare capable of accelerating particles to this momentum occurs approximately every 25 days with $E_{\rm flare}=3\times 10^{31}\,$erg (see Fig.\,\ref{fig:flare}). Considering the grey shaded region for the flare rate corresponds to a range of a flare every 7 or 72 days. 

While \citet{gunther_2020} suggest that the flare rates given by the dashed grey lines are likely to be associated with stars with $P_{\rm rot}<10$\,days (in comparison to GJ\,436 with $P_{\rm rot}\sim 44\,$days), it is interesting to note that the flare rate given by the solid green line is for an M dwarf with $P_{\rm rot}\sim 27\,$days. This rotation period is similar to the Sun's rotation period but with a flare rate 2 orders of magnitude larger than the Sun's. If GJ\,436 had such an elevated flare rate, flares capable of accelerating particles to 50\,MeV energies would occur approximately once per day. Fig.\,\ref{fig:pmax} \citep[and more broadly Fig.\,11 from][]{gunther_2020} also shows that the power law fit for the flare rate varies from star to star. Thus, determining stellar flare rates is important to understand how continuous stellar energetic particle fluxes of a certain energy are. Comparing flare rates with chemical recombination rates in the future would shed light on whether the chemistry in exoplanet atmospheres could reset between flares with associated stellar energetic particles.

\subsection{Connection to habitability}
Understanding exoplanet habitability, and detecting the signature of life as we know it from an exoplanet, is a key goal of modern astronomy. Energetic particles are one of the many factors thought to affect exoplanet habitability \citep[e.g.][]{meadows_2018}. For instance, energetic particles can drive the formation of prebiotic molecules \citep{airapetian_2016,barth_2021} that are important for the origin of life. On the other hand, high energetic particle fluxes may be detrimental to existing life by contributing to ozone depletion \citep{segura_2010} which blocks harmful UV radiation reaching the surface of an Earth-like planet. Studying exoplanets that are not expected to be habitable (such as GJ\,436\,b), but that will be observed in greater numbers, can provide important information about the conditions present in other stellar systems.

While hydrogen dominated gas giants are not generally thought to be conducive to life, \citet{Seager2021} have suggested how life might survive in the cloud decks of these planets, and \citet{Seager2020} have found unique biosignature gas compositions for microbial life as we know it in H$_2$-rich environments. Prebiotically relevant molecules, such as the amino acid glycine, may also form more easily on charged cloud particles \citep{2014IJAsB..13..165S}. \citet{globus_2020} suggested muons (which are on-average spin-polarised) that reach Earth's surface could be responsible for the chirality of the DNA helix by interacting with biological molecules. In this context, it will be of interest in the future to calculate the muon fluxes associated with our (proton) energetic particle fluxes for a terrestrial planet. While genetic mutation rates are known to increase with increasing radiation dose \citep{muller_1927}, it remains unclear what the ultimate effect of high radiation doses due to energetic particles would be on the development of life on another planet. Lifeforms that use radiation energy for metabolism \citep{matusiak_2016} may dominate instead.

In our model, we have assumed that the top-of-atmosphere stellar energetic particle fluxes are isotropic. In reality, stellar energetic particles will travel along spiral flux tubes which would result in anisotropic top-of-atmosphere fluxes. This type of transport can be modelled with the focused transport equation, for instance \citep{roelof_1969}. For tidally locked planets, if the stellar energetic particle flux is not isotropic, one side of the atmosphere, facing toward the star, will likely receive more stellar energetic particles than the side facing away from the star. This angular dependence could affect the energetic particle environment and could introduce heterogeneous chemistry in the upper atmosphere, depending on chemical timescales and efficiency of atmospheric circulation. This could be explored by 3D models, and is outside the scope of this paper. An anisotropic energetic particle flux could also have implications for habitability. Again, exoplanetary magnetic fields will complicate the picture further. For instance, \citet{herbst_2019} considered the impact of an Earth-like magnetic field and indicated that $<510$\,MeV energy particles were not able to penetrate an Earth-like magnetosphere at latitudes around 60$^\circ$.

The atmospheres of the planets considered here are roughly analogous to the most extreme transient post-impact atmospheres predicted for the early Earth \citep{zahnle_2020,itcovitz_2022}. We can consider the energetic particle flux at different pressures as analogous particle radiation environments for a post-impact early Earth. If Earth post-impact had a 100 bar ${\rm H_2}$ atmosphere, as proposed by \citet{zahnle_2020}, then Galactic cosmic rays determine the surface irradiation environment, and the irradiation dose is much lower than if Earth post-impact had a relatively more tenuous atmosphere, at $<10$ bar. Then the irradiation is much more severe and due to stellar energetic particles. This is relevant for habitability around small cool stars, such as GJ\,436, where the 0.2\,au case is at the inner edge of the liquid water habitable zone of the system. This is also relevant for prebiotic chemistry driven by energetic particle-driven chemistry in aqueous solutions, either on the surface or in aerosols \citep{Kobayashi2001,Lingam2018}.

\section{Conclusions}
\label{sec:conclusions}

In this paper, we consider the effect of the stellar magnetic environment and planetary atmosphere on the energy dependence and intensity of energetic particles as applied to the well-studied M dwarf system, GJ\,436. GJ\,436\,b is a warm Neptune planet orbiting at 0.028\,au from its host star (scheduled for JWST observations and an Ariel target). We have coupled a stellar wind model with an energetic particle transport model for the large-scale stellar system. We have then propagated the top-of-atmosphere energetic particle fluxes through the atmosphere of a warm Neptune-like exoplanet at various orbital distances from GJ\,436, using temperature pressure profiles from the radiative transport code HELIOS. 

We consider two sources of energetic particles: Galactic cosmic rays and flare-accelerated stellar energetic particles. We show how the maximum momentum that the stellar energetic particles are accelerated to can be related to the stellar X-ray luminosity. Thus, our input stellar energetic particle spectrum reflects in some sense the activity of the star. To our knowledge, no other model relates the cut-off energy of the stellar energetic particle spectrum to a stellar property. The high energy cut-off is very important because the highest energy particles (i.e. GeV energies) are those that can reach the surface of an Earth-like planet, for instance. Our calculations have assumed that the stellar energetic particle fluxes are constant in time up to our calculated maximum stellar energetic particle momentum. These would be associated with a $\sim10^{31}$\,erg flare which would occur once every $\sim$25 days assuming 10 times the solar flare rate for GJ\,436.

We calculate top-of-atmosphere energetic particle fluxes as a function of star-planet separation. We also model the transport of energetic particles through a hydrogen-dominated gas giant atmosphere at orbital distances of $a=0.01-0.2\,$au. The peak top-of-atmosphere ionisation rate for Galactic cosmic rays is $\sim 10^6$ times less than the peak ionisation rate for stellar energetic particles at orbital distances between $a=0.01-0.2\,$au. However, the stellar energetic particles do not penetrate as deep into the atmospheres. This is because for GJ\,436, the stellar energetic particle spectrum that we adopt has a lower cutoff energy than for the Galactic cosmic rays. 

At the orbital distance of GJ\,436\,b, we find that the ion-pair production rate, $Q$ (cm$^{-3}$\,s$^{-1}$), peaks at $\sim 100\,$bar for Galactic cosmic rays and $10^{-3}\,$bar for stellar energetic particles. We also estimate the skin-depth equivalent dose rate from the primary energetic particles as a function of height in the hydrogen-dominated atmospheres. Given the close-in orbital distance of the inner edge of GJ\,436's habitable zone (0.2\,au), the top-of-atmosphere equivalent dose rates that we calculate are substantial ($10^{4}\,$mSv\,day$^{-1}$). However, an important future step would be to study the effect of a planetary magnetic field on our results.

The stellar energetic particle fluxes decrease with distance from the star and we assume that they are isotropic at the top of the exoplanet atmosphere. At small orbital distances, the fluxes drop close to $1/r^2$ as advection processes dominate. At the same time, the magnetised, expanding stellar wind suppresses Galactic cosmic ray fluxes, and so the greater the distance from the star, the more intense the Galactic cosmic ray spectrum. A future step would be to study the effect of anistropic stellar energetic particle fluxes as they travel along magnetic flux tubes.

We find that, at a distance of 10\,au, the top-of-atmosphere molecular hydrogen ionisation rates from stellar energetic particles and Galactic cosmic rays are equal. Top-of-atmosphere ionisation by energetic particles is dominated by Galactic cosmic rays for $a>10\,$au. Ionisation by energetic particles at the top of the atmosphere is dominated by stellar energetic particles for $a<10\,$au. We anticipate that most of the ion chemistry will happen at $P< 1 \mu$bar, and so the ion chemistry will be determined by stellar energetic particles and UV photons for planets within 10\,au based on the top-of-atmosphere ionisation rates that we found. Clouds might also form in GJ\,436\,b's atmosphere because of the low atmospheric temperatures. Their formation and its effects on the atmospheric structure and the chemistry are, however, neglected in this paper. The chemical consequences of these trends and specific predictions for ion-neutral chemistry will be discussed in more detail in Paper II \citep{Rimmer2023}.

\section*{Acknowledgements}
DRL would like to acknowledge that this publication has emanated from research conducted with the financial support of Science Foundation Ireland under Grant number 21/PATH-S/9339. DRL, ALM and AAV acknowledge funding from the European Research Council (ERC) under the European Union's Horizon 2020 research and innovation programme (grant agreement No 817540, ASTROFLOW). DRL wishes to acknowledge the Irish Centre for High-End Computing (ICHEC) for the provision of computational facilities and support. Research reported in this publication was supported by the Royal Irish Academy. ChH acknowledges funding from the European Union H2020-MSCA-ITN-2019 under Grant Agreement no. 860470 (CHAMELEON). OV acknowledges funding from the ANR project `EXACT' (ANR-21-CE49-0008-01), from the Centre National d'\'{E}tudes Spatiales (CNES), and from the CNRS/INSU Programme National de Plan\'etologie (PNP). PB acknowledges a St Leonard’s Interdisciplinary Doctoral Scholarship from the University of St Andrews. We would like to thank the referee for helpful comments which improved the manuscript.

\section*{Data Availability}
The output data underlying this article will be available via zenodo.org upon publication.

\bibliographystyle{mnras}
\bibliography{bibfile}

\appendix
\section{Flare rates}
\begin{figure}
	\centering
        \includegraphics[width=0.5\textwidth]{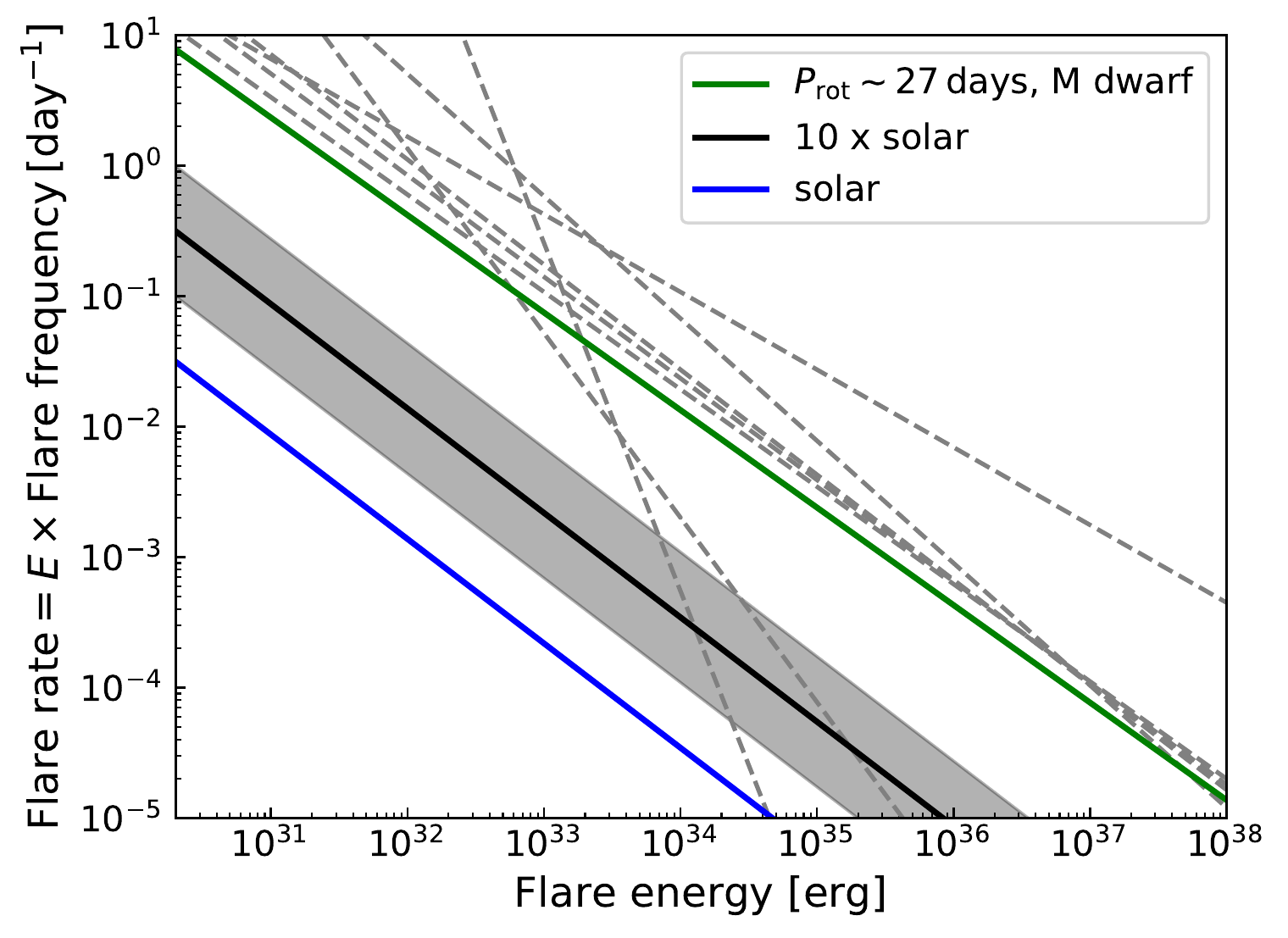}
       	\centering
  \caption{Observationally inferred flare rates ($E_{\rm flare}dN/dE_{\rm flare}$) are plotted as a function of flare energy for the Sun, 10 times the solar flare rate and a number of M dwarf flare rates from \citet{gunther_2020}. The linestyles and the grey shaded region are the same as for Fig.\,\ref{fig:pmax}. \label{fig:flare}} 
\end{figure}

Section\,\ref{subsec:continuous} discusses how continuous stellar energetic particle fluxes can be considered for GJ\,436. Fig.\,\ref{fig:pmax} relates the flare rate to a maximum stellar energetic particle momentum. This is based on the underlying relation between flare rate and flare energy which is plotted in Fig.\,\ref{fig:flare}. The solar flare rate is a fit to solar and stellar super-flare data from \citet{maehara_2015} with a flare frequency relation of $dN/dE_{\rm flare} \propto E_{\rm flare}^{-1.8}$. The grey shaded region surrounding the solid black line, representing 10 times the solar flare rate, is $\pm 0.5\,$dex. While the flare rate for GJ\,436 is unknown, given the maximum stellar energetic particle momentum that we assume, we indicated the frequency of flares accelerating to this momentum in Section\,\ref{subsec:continuous}.

\label{lastpage}

\end{document}